\documentclass{article}

    \usepackage[final,nonatbib]{neurips_2024}

\usepackage[utf8]{inputenc} 
\usepackage[T1]{fontenc}    
\usepackage{hyperref}       
\usepackage{url}            
\usepackage{booktabs}       
\usepackage{amsfonts}       
\usepackage{nicefrac}       
\usepackage{microtype}      
\usepackage{xcolor}         
\usepackage{amsmath}
\usepackage{amsthm}
\usepackage{amssymb}
\usepackage{graphicx}
\usepackage{algorithm}

\usepackage[noend]{algpseudocode}
\usepackage{caption}
\usepackage{subcaption}
\usepackage{multirow}
\usepackage{xspace}
\usepackage{multicol}

\newtheorem{theorem}{Theorem}
\newtheorem{lemma}{Lemma}
\newtheorem{problem}{Problem}
\newtheorem{proposition}{Proposition}

\newcommand{\sysname}{{SEEV}\xspace}

\title{\sysname: Synthesis with Efficient Exact Verification for ReLU Neural Barrier Functions}

\author{Hongchao Zhang\thanks{Equal contribution}\\
    Electrical \& Systems Engineering\\
    Washington University in St. Louis \\ 
    \texttt{hongchao@wustl.edu}
    \And
    Zhizhen Qin\footnotemark[1]\\
    Computer Science \& Engineering\\
    University of California, San Diego \\ 
    \texttt{zhizhenqin@ucsd.edu}
    \And
    Sicun Gao \\
    Computer Science \& Engineering\\
    University of California, San Diego \\ 
    \texttt{scungao@ucsd.edu}
    \And
    Andrew Clark \\
    Electrical \& Systems Engineering\\
    Washington University in St. Louis \\
    \texttt{andrewclark@wustl.edu}
}

\begin{document}
\maketitle
\begin{abstract}
Neural Control Barrier Functions (NCBFs) have shown significant promise in enforcing safety constraints on nonlinear autonomous systems. State-of-the-art exact approaches to verifying safety of NCBF-based controllers exploit the piecewise-linear structure of ReLU neural networks, however, such approaches still rely on enumerating all of the activation regions of the network near the safety boundary, thus incurring high computation cost. In this paper, we propose a framework for Synthesis with Efficient Exact Verification (SEEV). Our framework consists of two components, namely (i) an NCBF synthesis algorithm that introduces a novel regularizer to reduce the number of activation regions at the safety boundary, and (ii) a verification algorithm that exploits tight over-approximations of the safety conditions to reduce the cost of verifying each piecewise-linear segment. Our simulations show that SEEV significantly improves verification efficiency while maintaining the CBF quality across various benchmark systems and neural network structures. Our code is available at \url{https://github.com/HongchaoZhang-HZ/SEEV}. 

\end{abstract}

\section{Introduction}
\label{sec:intro}
Safety is a crucial property for autonomous systems that interact with humans and critical infrastructures in applications including medicine, energy, and robotics~\cite{hsu2015control,agrawal2021safe}, which has motivated recent research into safe control~\cite{xu2017correctness,breeden2021high, dai2023convex, kang2023verification,rober2023hybrid}. Control Barrier Functions (CBFs), which apply a constraint on the control input at each time in order to ensure that safety constraints are not violated, have attracted significant research attention due to their ease of implementation and compatibility with a variety of safety and performance criteria \cite{dawson2023safe}. Recently, CBFs that are defined by neural networks, denoted as Neural Control Barrier Functions (NCBFs), have been proposed to leverage the expressiveness of NNs for safe control of nonlinear systems \cite{so2023train,dawson2022safe,tayal2024learning}. NCBFs have shown substantial promise in applications including robotic manipulation \cite{dawson2022safe}, navigation \cite{long2021learning,xiao2023barriernet}, and flight control \cite{zhao2021learning}.

A key challenge in NCBF-based control is safety verification, which amounts to ensuring that the constraints on the control can be satisfied throughout the state space under actuation limits. 
The NCBF safety verification problem effectively combines two problems that are known to be difficult, namely, input-output verification of neural networks (VNN) \cite{zhang2018efficient,xu2020automatic,salman2019convex,xu2021fast,wang2021beta,zhang22babattack} and reachability verification of nonlinear systems. 
While sound and complete verifiers such as dReal can be applied to NCBFs, they typically can only handle systems of dimension three or small neural networks \cite{abate2021fossil, edwards2023fossil}. In \cite{zhang2023exact}, exact conditions for safety verification of NCBFs with ReLU activation functions were proposed that leverage the piecewise-linearity of ReLU-NNs to reduce verification time compared to dReal for general activation functions. The exact conditions, however, still require checking correctness of the NCBF by solving a nonlinear optimization problem along each piecewise-linear segment.  Hence, the NCBF verification problem remains intractable for high-dimensional systems.

In this paper, we propose a framework for Synthesis with Efficient Exact Verification (SEEV) for piecewise-linear NCBFs. The main insight of SEEV is that the computational bottleneck of NCBF verification is the inherent requirement of verifying each linear segment of the neural network. We mitigate this bottleneck by (i) developing a training procedure that reduces the number of segments that must be verified and (ii) constructing verification algorithms that efficiently enumerate the linear segments at the safety boundary and exploit easily-checked sufficient conditions to reduce computation time. Towards (i), we introduce a regularizer to the loss function that penalizes the dissimilarity of activation patterns along the CBF boundary. Towards (ii), we propose a breadth-first search algorithm for efficiently enumerating the boundary segments, as well as tight linear over-approximations of the nonlinear optimization problems for verifying each segment. Moreover, we integrate the synthesis and verification components by incorporating safety counterexamples returned by the safety verifier into the training dataset. Our simulation results demonstrate significant improvements in verification efficiency and reliability across a range of benchmark systems.

\textbf{Related Work: }
Neural control barrier functions have been proposed to describe complex safety sets to remain inside and certify safety of a controlled system ~\cite{berkenkamp2017safe,abate2021fossil,qin2021learning,qin2022quantifying} or synthesize control input based on NCBFs to ensure safety~\cite{dawson2023safe,so2023train,dawson2022safe,liu2023safe}. However, the synthesized NCBF may not ensure safety. Safety verification of NCBFs is required. 
Sum-of-squares (SOS) optimization~\cite{prajna2007framework, ames2019control, zhao2023safety, schneeberger2023sos, clark2021verification} has been widely used for polynomial barrier functions, however, they are not applicable due to the non-polynomial and potentially non-differentiable activation functions of NCBFs. 
VNN~\cite{ferrari2022complete, henriksen2021deepsplit, zhang2022general} and methods for ReLU neural networks~\cite{katz2017reluplex, katz2019marabou} are also not directly applicable to NCBF verification.
Nonlinear programming approach~\cite{zhang2023exact} provides another route for exact verification but is computationally intensive and relies on VNN tools.
%
To synthesize neural networks with verifiable guarantees, Counterexample Guided Inductive Synthesis (CEGIS) has been applied using SMT-based techniques~\cite{abate2021fossil,zhao2020synthesizing, abate2020formal, 10.1007/978-3-030-72016-2_20, anand2023formally}.
Other verification-in-the-loop approaches utilize reachability analysis~\cite{wang2022design} and branch-and-bound neural network verification tools~\cite{wang2024simultaneous}. However, existing works suffer from the difficulty of performing verification and generating counterexamples in a computationally efficient manner.
Sampling-based approaches~\cite{anand2023formally, tayal2024learning} aim to prove safety using Lipschitz conditions, but they rely on dense sampling over the state space, which is computationally prohibitive. 
In this work, we present \sysname to integrate the synthesis
and efficient verification by incorporating safety counterexamples from the exact verification.

\textbf{Organization} The remainder of the paper is organized as follows. Section \ref{sec:preliminaries} gives the system model and background on neural networks and the conditions of valid NCBFs. Section \ref{sec:synthesis} presents the SEEV framework. Section \ref{sec:verification} presents our efficient and exact verification. Section \ref{sec:experiments} contains simulation results. Section \ref{sec:conclusions} concludes the paper. 

\section{Preliminaries}
\label{sec:preliminaries}

This section presents the system model, notations on neural networks, and exact conditions of safety. 



\subsection{System Model}
\label{subsec:system-model}
We consider a system with state $x(t) \in \mathcal{X} \subseteq \mathbb{R}^{n}$ and input $u(t) \in \mathcal{U} \subseteq \mathbb{R}^{m}$, with initial condition $x(t_0)=x_0$ where $x_0$ lies in an initial set $\mathcal{I}\subseteq \mathcal{X}$. The continuous-time nonlinear control-affine system has the dynamics given by 
\begin{equation}
    \label{eq:dynamics}
    \dot{x}(t) = f(x(t)) + g(x(t))u(t)
\end{equation}
where $f: \mathbb{R}^{n} \rightarrow \mathbb{R}^{n}$ and $g: \mathbb{R}^{n} \rightarrow \mathbb{R}^{n \times m}$ are known continuous functions. 

We consider the case that the system is required to remain inside a given set of states, i.e., $x(t) \in \mathcal{C}$ for all time $t \geq t_{0}$. The set $\mathcal{C}$, referred to as the \emph{safe set}, is defined as $\mathcal{C} = \{x : h(x) \geq 0\}$ by some given continuous function $h: \mathbb{R}^{n} \rightarrow \mathbb{R}$. The unsafe region is given by  $\mathcal{X}\setminus\mathcal{C}$.  

\subsection{Neural Network Model and Notations}
\label{subsec:neural-network}
We let $\mathbf{W}$ and $\mathbf{r}$ denote the weight and bias of a neural network, and let $\theta$ be a parameter vector obtained by concatenating $\mathbf{W}$ and $\mathbf{r}$. We consider a $\theta$-parameterized feedforward neural network $b_{\theta}: \mathbb{R}^{n} \rightarrow \mathbb{R}$ constructed as follows. The network consists of $L$ layers, with each layer $i$ consisting of $M_{i}$ neurons. We let $(i,j) \in \{1,\ldots,L\} \times \{1,\ldots,M_{i}\}$ denote the $j$-th neuron at the $i$-th layer. We denote the pre-activation input as $z_{j}^{(i)}$, piecewise linear activation function $\sigma$ and the post-activation output as $\hat{z}_{j}^{(i)} = \sigma(z_{j}^{(i)})$. Specifically, we assume that the NN has the Rectified Linear Unit (ReLU) activation function, defined by $\sigma(z) = z$ for $z \geq 0$ and $\sigma(z)  = 0$ for $z < 0$. We define the neuron $(i,j)$ as \emph{active} if $z_{j}^{(i)}>0$, \emph{inactive} if $z_{j}^{(i)}<0$ and \emph{unstable} if $z_{j}^{(i)}=0$. Let $\mathbf{S} = \tau_S(x) = \{(i,j) : z_{j}^{(i)} \geq 0\} \subseteq \{(i,j) : i=1,\ldots,L, j=1,\ldots,M_{i}\}$ denote the set of activated and unstable neurons, produced by state $x$ and function $\tau_S$. Let $\mathbf{T}(x) = \tau_T(x) = \{(i,j) : z_{j}^{(i)} = 0\}$ denote the set of unstable neurons produced by $x$ and $\tau_T$. $\mathbf{T}(\mathbf{S}_1, \ldots, \mathbf{S}_r)$ denote the set of unstable neurons produced by activation sets $\mathbf{S}_1, \ldots, \mathbf{S}_r$. The set of inactive neurons is given by $\mathbf{S}^{c}$, i.e., the complement of $\mathbf{S}$, and consists of neurons with negative pre-activation input.
We define vectors $\overline{W}_{ij}(\mathbf{S}) \in \mathbb{R}^{n}$ and scalars $\overline{r}_{ij}(\mathbf{S}) \in \mathbb{R}$ such that $z_{j}^{(i)} = \overline{W}_{ij}(\mathbf{S})^{T}x + \overline{r}_{ij}(\mathbf{S})$ in the Appendix \ref{appendix:prelim}.
The symmetric difference between two sets $\mathbf{A}$ and $\mathbf{B}$, denoted by $\mathbf{A} \Delta \mathbf{B}$, is defined as 
$\mathbf{A} \Delta \mathbf{B} = (\mathbf{A} \setminus \mathbf{B}) \cup (\mathbf{B} \setminus \mathbf{A})$. 

Finally, we define the terms \emph{hyperplane} and \emph{hinge}. For any $\mathbf{S}\subseteq \{1,\ldots,L\} \times \{1,\ldots,M_{i}\}$, we define $\overline{\mathcal{X}}(\mathbf{S}):=\{x: \mathbf{S}(x)=\mathbf{S}\}$. The collection of $\overline{\mathcal{X}}(\mathbf{S})$ for all $\mathbf{S}$ is the set of \emph{hyperplanes} associated with the ReLU neural network. A hyperplane that intersects the set $\{x: b_{\theta}(x) = 0\}$ is a \emph{boundary hyperplane}.
The intersection of hyperplanes $\overline{\mathcal{X}}(\mathbf{S}_{1}),\ldots,\overline{\mathcal{X}}(\mathbf{S}_{r})$ is called a \emph{hinge}. A hinge that intersects the set $\{x: b_{\theta}(x) = 0\}$ is a \emph{boundary hinge}. 

\subsection{Guaranteeing Safety via Control Barrier Functions}
\label{subsec:safetyguarantee}


Barrier certificates \cite{prajna2007framework} ensure the safety of a feedback-controlled system under policy $\mu(x\mid \lambda)$ by identifying a CBF to represent the invariant safe set. The barrier certificate defines an inner safe region $\mathcal{D} := \{x :b(x) \geq 0\}$ for some continuous function $b$. The verifiable invariance of $\mathcal{D}$ is obtained from the following result.

\begin{theorem}[Nagumo's Theorem \cite{blanchini2008set}, Section 4.2]
\label{theorem:Nagumo}
A closed set $\mathcal{D}$ is controlled positive invariant if, whenever $x \in \partial \mathcal{D}$, where $\partial \mathcal{D}$ denotes the boundary of $\mathcal{D}$. we have 
\begin{equation}
\label{eq:Nagumo-CPI}
(f(x) + g(x)u) \in \mathcal{A}_{\mathcal{D}}(x)
\end{equation}
for some $u \in \mathcal{U}$ 
where $\mathcal{A}_{\mathcal{D}}(x)$ is the tangent cone to $\mathcal{D}$ at $x$.
\end{theorem}
We denote a state $\hat{x}_{ce}^{c}$ with $\hat{x}_{ce}^{c} \in \partial \mathcal{D}$ that violates \eqref{eq:Nagumo-CPI} as a \emph{safety counterexample}.
In the case where $b$ is continuously differentiable, \eqref{eq:Nagumo-CPI} can be satisfied by selecting $u$ to satisfy the condition 
$\frac{\partial b}{\partial x}(f(x(t)) + g(x(t))u(t)) \geq -\alpha(b(x(t)))$,
where $\alpha: \mathbb{R} \rightarrow \mathbb{R}$ is a strictly increasing function with $\alpha(0) = 0$. When $b$ is not continuously differentiable, as in a ReLU NCBFs, a modified condition is needed. 
Prior work \cite{zhang2023exact} introduces exact conditions for safety verification of ReLU NCBFs, based on the following proposition. 
A collection of activation sets $\mathbf{S_1},\ldots,\mathbf{S_r}$ is complete if, for any $\mathbf{S}^{\prime} \notin \{\mathbf{S_1},\ldots,\mathbf{S_r}\}$, we have $\overline{\mathcal{X}}(\mathbf{S_1}) \cap \cdots \cap \overline{\mathcal{X}}(\mathbf{S_r}) \cap \overline{\mathcal{X}}(\mathbf{S}^{\prime}) = \emptyset$.
\begin{proposition}
\label{prop:safety-condition}
Suppose the function ReLU neural network-defined function $b$ satisfies the following conditions:
\begin{enumerate}
\item[(i)] For all activation sets $\mathbf{S}_{1},\ldots,\mathbf{S}_{r}$ with $\{\mathbf{S}_{1},\ldots,\mathbf{S}_{r}\}$ complete 
 and any $x$ satisfying $b(x) = 0$ and 
\begin{equation}
\label{eq:safety-condition-x}
x \in \left(\bigcap_{l=1}^{r}{\overline{\mathcal{X}}(\mathbf{S}_{l})}\right),
\end{equation}
there exist $l \in \{1,\ldots,r\}$ and $u \in \mathcal{U}$ such that 
\begin{align}
\label{eq:safety-condition-1}
(\overline{\mathbf{W}}_{i-1}(\mathbf{S}_{l})W_{ij})^{T}(f(x) + g(x)u) &\geq  0 \quad \forall (i,j) \in \mathbf{T}(\mathbf{S}_{1},\ldots,\mathbf{S}_{r}) \cap \mathbf{S}_{l} \\
\label{eq:safety-condition-1a}
(\overline{\mathbf{W}}_{i-1}(\mathbf{S}_{l})W_{ij})^{T}(f(x) + g(x)u) &\leq  0 \quad \forall (i,j) \in \mathbf{T}(\mathbf{S}_{1},\ldots,\mathbf{S}_{r}) \setminus \mathbf{S}_{l} \\
\label{eq:safety-condition-2}
\overline{W}(\mathbf{S}_{l})^{T}(f(x) + g(x)u) &\geq 0 
\end{align}
\item[(ii)] For all activation sets $\mathbf{S}$, we have 
\begin{equation}
\label{eq:neural-containment}
(\overline{\mathcal{X}}(\mathbf{S}) \cap \mathcal{D}) \setminus \mathcal{C} = \emptyset
\end{equation}
\end{enumerate}
If $b(x(0)) \geq 0$, then $x(t) \in \mathcal{C}$ for all $t \geq 0$.
\end{proposition}

Any feedback control law $\mu: \mathcal{X} \rightarrow \mathcal{U}$ that satisfies \eqref{eq:safety-condition-1}--\eqref{eq:safety-condition-2} is guaranteed to ensure safety and is referred to as an NCBF control policy. 
Given a nominal control policy $\pi_{nom}(x)$, safe actions can be derived from a ReLU NCBF as a safety filter\cite{ames2019control,so2023train} by solving the following optimization problem proposed in \cite[Lemma 2]{zhang2023exact}:
\begin{equation}
\label{eq:safety filter}
    \min_{\mathbf{S} \in \mathbf{S}(x), u}  ||u-\pi_{nom}(x)||_{2}^{2} \quad
    \mbox{s.t.} \quad
    \overline{W}(\mathbf{S})^{T}(f(x) + g(x)u) \geq -\alpha(b(x))
    , u \in \mathcal{U},   (\ref{eq:safety-condition-1})- (\ref{eq:safety-condition-2}) 
\end{equation}
The solution to this optimization problem provides a control $u$ that minimally deviates from the nominal control $\pi_{nom}(x)$ while satisfying NCBF constraints derived in Proposition \ref{prop:safety-condition} ensuring that $\mathcal{D}$ is positive invariant and is contained in $\mathcal{C}$. 
Based on Proposition \ref{prop:safety-condition}, we can define different types of safety counterexamples. Correctness counterexamples, denoted by $\hat{x}_{ce}^{c}$, refers to a state $\hat{x}_{ce}^{c} \in \mathcal{D} \cap (\mathcal{X} \setminus \mathcal{C})$. Hyperplane verification counterexamples refer to states $\hat{x}_{ce}^{h}$ that violate \eqref{eq:safety-condition-2}. Hinge verification counterexamples are states $x$ with $\mathbf{T}(x) \neq \emptyset$ that violate \eqref{eq:safety-condition-1}--\eqref{eq:safety-condition-2}.


\section{Synthesis}
\label{sec:synthesis}




In this section, we present the framework to synthesize NCBF $b_{\theta}(x)$ to ensure the safety of the system \eqref{eq:dynamics}. The synthesis framework aims to train an NCBF and construct an NCBF-based safe control policy. We first formulate the problem and present an overview of the framework in \ref{subsec:synth_overview}. Then we demonstrate the design of the loss function in \ref{subsec:lossdesign}. 

\subsection{Overall Formulation}
\label{subsec:synth_overview}

    
Our primary objective is to synthesize a ReLU Neural Control Barrier Function (ReLU-NCBF) for \eqref{eq:dynamics} and develop a safe control policy to ensure system safety.

\begin{problem}
\label{prob:synth}
    Given a system \eqref{eq:dynamics}, initial set $\mathcal{I}$ and a safety set $\mathcal{C}$, synthesize a ReLU NCBF $b_{\theta}(x)$ parameterized by $\theta$ such that the conditions in Proposition \ref{prop:safety-condition} are satisfied. Then construct a control policy to synthesize $u_{t}$ such that the system remains positive invariant in $\mathcal{D}:=\{x: b_{\theta}(x)\geq0\}\subseteq \mathcal{C}$. 
\end{problem}

\begin{figure}
    \centering
    \includegraphics[width=0.75\textwidth]{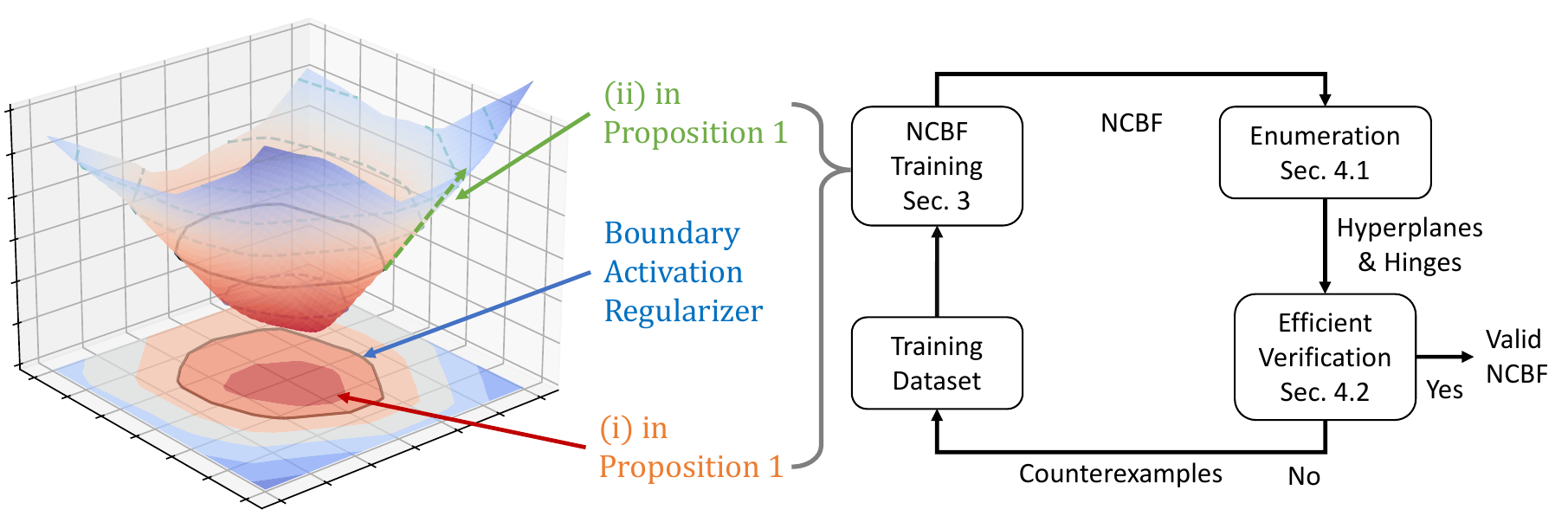}
    \caption{\sysname: Synthesis with Efficient Exact Verifier for ReLU NCBF}
    \label{fig:synthframe}
\end{figure}

We propose \sysname to address this problem with the synthesis framework demonstrated in Fig. \ref{fig:synthframe}. The training dataset $\mathcal{T}$ is initialized by uniform sampling over $\mathcal{X}$. The training framework consists of two loops. The inner loop attempts to choose parameter $\theta$ for $b_\theta(x)$ to satisfy the safety condition by minimizing the loss function over training data $\mathcal{T}$. 
The outer loop validates a given NCBF $b_\theta(x)$ by searching for safety counterexamples $\hat{x}_{ce}$ and updates the training dataset as $\mathcal{T}\cup \{\hat{x}_{ce}\}$. 


To train the parameters of the NCBF to satisfy the conditions of Proposition \ref{prop:safety-condition}, we propose a loss function that penalizes the NCBF for violating constraints (i) and (ii) at a collection of sample points. The loss function is a weighted sum of three terms. The first term is the correctness loss penalizing state $\hat{x} \in \mathcal{X}\setminus\mathcal{C}$ with $b_\theta(x)\geq 0$. The second term is verification loss that penalizes states $\hat{x}$ that $\nexists u$ such that \eqref{eq:safety-condition-1}-\eqref{eq:safety-condition-2} hold. The third term is a regularizer minimizing the number of hyperplanes and hinges along the boundary. 
However, minimizing the loss function is insufficient to ensure safety \cite{so2023train} because there may exist safety counterexamples outside of the training dataset. 
In order to guarantee safety, \sysname introduces an efficient exact verifier to certify whether $b_{\theta}(x)$ meets the safety conditions outlined in Proposition \ref{prop:safety-condition}. The verifier either produces a proof of safety or generates a safety counterexample that can be added to the training dataset to improve the NCBF. 

The integration of the verifier can improve safety by adding counterexamples to guide the training process, however, it may also introduce  additional computation complexity. 
We propose a combined approach, leveraging two complementary methods to address this issue. First, the verification of \sysname introduces an efficient algorithm in Section \ref{sec:verification} to mitigate the computational scalability challenges that arise as neural network depth and dimensionality increase. Second, \sysname introduces a regularizer to limit the number of boundary hyperplanes and hinges to be verified, addressing the complexity that arises as neural network size increases.

\subsection{Loss Function Design and NCBF Training}
\label{subsec:lossdesign}

The goal of the inner loop is to choose parameters $\theta$ so that the conditions of Proposition \ref{prop:safety-condition} are satisfied for all $\hat{x}\in \mathcal{T}$ and the computational cost of verifying safety is minimized. To achieve the latter objective, we observe (based on results presented in Table \ref{tab:NCBFcomparison}) that the computational complexity of verification grows with the cardinality of the collection of activation sets that intersect the safety boundary $\partial\mathcal{D}$. The collection is defined as $\mathcal{B}:=\{\mathbf{S}_i: \partial\mathcal{D} \cap \overline{\mathcal{X}}(\mathbf{S}_i)\neq \emptyset\}$. 
Hence, we formulate the following unconstrained optimization problem to search for $\theta$. 
\begin{equation}
    \label{eq:uncons_opt}
    \min_{\theta}\quad{\lambda_\mathcal{B}\mathcal{L}_\mathcal{B}(\mathcal{T})} + \lambda_{f}\mathcal{L}_f(\mathcal{T}) + \lambda_{c}\mathcal{L}_c(\mathcal{T}) 
\end{equation}
where $\mathcal{L}_\mathcal{B}(\mathcal{T})$ regularizer to minimize $|\mathcal{B}|$, $\mathcal{L}_f(\mathcal{T})$ is the loss penalizing the violations of constraint \eqref{eq:safety-condition-1}-\eqref{eq:safety-condition-2} ((i) of Proposition \ref{prop:safety-condition}), $\mathcal{L}_c(\mathcal{T})$ penalizes the violations of constraint \eqref{eq:neural-containment} ((ii) of Proposition \ref{prop:safety-condition}), and $\lambda_\mathcal{B}$, $\lambda_{f}$ and $\lambda_{c}$ are non-negative coefficients 
defined as follows.

\textbf{$\mathcal{L}_f$ Regularizer:}
For each sample $\hat{x}\in \mathcal{T}$ the safe control signal is calculated by 
\begin{equation}
\label{eq:nominal_controller}
    \min_{u,r} \quad ||u-\pi_{nom}(x)||_{2}^{2} \qquad
    s.t. \quad \mathcal{W}(\mathbf{S}_{l})^{T}(f(\hat{x}) + g(\hat{x})u) + r \geq 0 
\end{equation}
where $\mathcal{W} = \overline{W}$ for differentiable points and $\mathcal{W} = \tilde{W}$ defined as the subgradient at non-differentiable points. The regularizer $\mathcal{L}_f$ enforces the satisfaction of the constraint by inserting a positive relaxation term $r$ in the constraint and minimizing $r$ with a large penalization in the objective function. 
We have the loss $\mathcal{L}_f$ defined as $\mathcal{L}_f = ||u-\pi_{nom}(x)||_{2}^{2} + r$.
We use \cite{cvxpylayers2019} to make this procedure differentiable, allowing us to employ the relaxation loss into the NCBF loss function design.

\textbf{$\mathcal{L}_c$ Regularizer:}
$\mathcal{L}_c$ regularizer enforces the correctness of the NCBF. In particular, it enforces the $b_{\theta}(x)$ of safe samples $x \in \mathcal{T}_{\mathcal{I}}$ to be positive, and $b_{\theta}(x)$ of unsafe samples $\mathcal{T}_{\mathcal{X}\setminus\mathcal{C}}$ to be negative. Define $N_{\text{safe}}=|\mathcal{T}_{\mathcal{I}}|$ and $N_{\text{unsafe}}=|\mathcal{T}_{\mathcal{X}\setminus\mathcal{C}}|$, and with a small positive tuning parameter $\epsilon > 0$, the loss term $\mathcal{L}_c$ can be defined as
\begin{equation}
\label{eq:corr_loss}
    \mathcal{L}_c = a_1 \frac{1}{N_{\text{safe}}} \sum_{x \in \mathcal{T}_{\mathcal{I}}} [\epsilon - b_{\theta} (x) ]_{+} + a_2 \frac{1}{N_{\text{unsafe}}} \sum_{x \in \mathcal{T}_{\mathcal{X}\setminus\mathcal{C}}} [\epsilon + b_{\theta} (x)]_{+}
\end{equation}
where $[\cdot]_{+}=max(\cdot, 0)$. $a_1$ and $a_2$ are positive parameters controlling penalization strength of the violations of safe and unsafe samples.

\textbf{$\mathcal{L}_\mathcal{B}$ Regularizer:} We propose a novel regularizer to limit the number of boundary hyperplanes and hinges by penalizing the dissimilarity, i.e., $\mathbf{S}(\hat{x}_i)\Delta\mathbf{S}(\hat{x}_j)$ of boundary activation sets $\mathbf{S}(\hat{x})\in \mathcal{B}$. 
However, the dissimilarity measure of boundary activation sets is inherently nondifferentiable. To address this issue the regularizer introduces the generalized sigmoid function $\sigma_k(z) = \frac{1}{1 + \exp(-k \cdot z)}$ to compute the vector of smoothed activation defined as $\phi_{\sigma_k}(x):=[\sigma_k(z_{i,j}), \forall i,j\in \{1,\ldots,L\} \times \{1,\ldots,M_{i}\}]$. 
The $\mathcal{L}_\mathcal{B}$ regularizer conducts the following two steps to penalize dissimilarity. 

In the first step, the regularizer identifies the training data in the boundary hyperplanes and hinges denoted as $\hat{x}\in \mathcal{T}_{\partial\mathcal{D}}$. The set is defined as $\mathcal{T}_{\partial\mathcal{D}}:=\{\hat{x}:\hat{x}\in \overline{\mathcal{X}}(\mathbf{S}), \forall \mathbf{S}\in \mathcal{B}\}$. 
To further improve the efficiency, the regularizer approximates $\mathcal{T}_{\partial\mathcal{D}}$ with a range-based threshold $\epsilon$ on the output of the NCBF, i.e.,  $|b_{\theta}(\hat{x})|\leq \epsilon$. 

The second step is to penalize the dissimilarity of $\mathbf{S}\in \mathcal{B}$. To avoid the potential pitfalls of enforcing similarity across the entire boundary $\mathcal{B}$, the regularizer employs an unsupervised learning approach to group the training data into $N_{\mathbf{B}}$ clusters. We define the collection of the activation set in each cluster as $\mathbf{B}_i\subseteq \mathcal{B}$ and the collection in each cluster as $\mathcal{T}_{\mathbf{B}_i}:=\{ \hat{x}: \hat{x}\in \bigcup_{\mathbf{S}\in \mathbf{B}_i} \overline{\mathcal{X}}(\mathbf{S}) \}$. 
The $\mathcal{L}_\mathcal{B}$ is then defined as follows, with an inner sum over all pairs of $\hat{x}\in \mathcal{T}_{\mathbf{B}_i}$ and an outer sum over all clusters. 
\begin{equation}
    \label{eq:loss_B}
    \mathcal{L}_{\mathcal{B}} = \frac{1}{N_{\mathbf{B}}} \sum_{\mathbf{B}_i \in \mathcal{B}} \frac{1}{|\mathcal{T}_{\mathbf{B}_i}|^2} \sum_{\hat{x}_{i},\hat{x}_{j}\in\mathcal{T}_{\mathbf{B}_i}} \| \phi_{\sigma_k}(x_i) - \phi_{\sigma_k}(x_j) \|_2^2,
\end{equation}

\section{Verification}
\label{sec:verification}
\begin{figure}[h]
    \centering
    \includegraphics[width=0.8\textwidth]{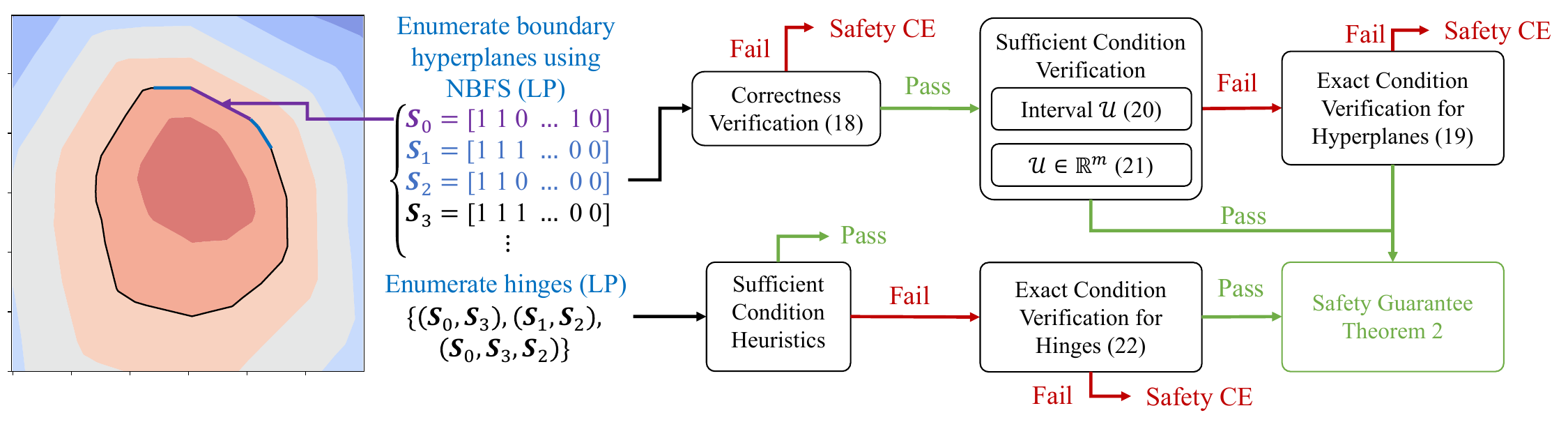}
    \caption{Overview of the Efficient Exact Verifier for ReLU NCBFs}
    \label{fig:sec4diagram}
\end{figure}

In this section, we demonstrate the efficient exact verification of the given NCBF $b_{\theta}(x)$ to ensure the positive invariance of the set $\mathcal{D}$ under the NCBF-based control policy. 
In what follows, we propose an efficient enumeration and progressive verification moving from sufficient to exact conditions. 
The overview of the proposed approach is as shown in Fig. \ref{fig:sec4diagram}. \sysname decomposes the NCBF into hyperplanes and hinges and verifies each component hierarchically, with novel tractable sufficient conditions verified first and the more complex exact conditions checked only if the sufficient conditions fail. 
Given an NCBF $b_{\theta}(x)$, the verification of \sysname returns (i) a Boolean variable that is `true' if the conditions of Proposition \ref{prop:safety-condition} are satisfied and `false' otherwise, and (ii) a safety counterexample $\hat{x}$ that violates the conditions of Proposition \ref{prop:safety-condition} if the result is `false'. 
 Algorithm \ref{alg:EEV} presents an overview of the verification of \sysname.
The algorithm consists of an enumeration stage to identify all boundary hyperplanes and hinges, and a verification stage to certify satisfaction of conditions in Proposition \ref{prop:safety-condition} for all hyperplanes and hinges.
\begin{algorithm}[h]
\caption{Efficient Exact Verification}
\begin{algorithmic}[1]
    \State \textbf{Input:} $n, \mathcal{T}_{\mathcal{X}\setminus\mathcal{C}}, \mathcal{T}_\mathcal{I}$
    \State \textbf{Output:} Verification Boolean result $r$, Categorized counterexample $\hat{x}_{ce}$
    \Procedure{Efficient Exact Verification}{$\mathcal{T}_{\mathcal{X}\setminus\mathcal{C}}, \mathcal{T}_\mathcal{I}$}
    \State $\mathbf{S}_0 \gets \text{ENUMINIT}(\mathcal{T}_{\mathcal{X}\setminus\mathcal{C}}, \mathcal{T}_\mathcal{I})$
    \Comment{Initial Activation Set Identification, Section \ref{subsec:enumeration}}
    \State $\mathcal{S} \gets \text{NBFS}(\mathbf{S}_0)$
    \Comment{Activation Sets Enumeration, Section \ref{subsec:enumeration}}
    \State $r,\hat{x}_{ce}^{(c)}\gets\text{CorrectnessVerifier}(\mathcal{S})$
    \Comment{Correctness Verification, Section \ref{subsec:efficient_veri}}
    \State $r,\hat{x}_{ce}^{(h)}\gets\text{HyperplaneVerifier}(\mathcal{S})$
    \Comment{Hyperplane Verification, Section \ref{subsec:efficient_veri}}
    \State $\mathcal{V} \gets \text{HingeEnum}(\mathcal{S}, n)$
    \Comment{Hinges Enumeration, Section \ref{subsec:enumeration}}
    \State $r,\hat{x}_{ce}^{(g)}\gets\text{HingeVerifier}(\mathcal{V})$
    \Comment{Hinge Verification, Section \ref{subsec:efficient_veri}}
    \State \textbf{Return} $r$, $\hat{x}_{ce}$
   \EndProcedure
\end{algorithmic}
\label{alg:EEV}
\end{algorithm}


\subsection{Enumeration of Hyperplanes and Hinges}
\label{subsec:enumeration}
The boundary enumeration identifies the boundary hyperplanes $\mathbf{S}_0\in \mathcal{B}$. It initially identifies the initial boundary activation set $\mathbf{S}_0\in \mathcal{B}$, enumerates all $\mathbf{S}\in \mathcal{B}$ by NBFS starting from $\mathbf{S}_0$, and finally enumerates all hinges consisting of the intersections of hyperplanes. The NBFS approach avoids over-approximation of the set of boundary hyperplanes that may be introduced by, e.g., interval propagation methods, and hence is particularly suited to deep neural networks.

In what follows, we assume that the unsafe region $\mathcal{X} \setminus \mathcal{C}$ and initial safe set $\mathcal{I}$ are connected, and use $\partial \mathcal{D}$ to refer to the connected component of the boundary of $\mathcal{D}$ that separates $\mathcal{I}$ and $\mathcal{X} \setminus \mathcal{C}$. This assumption is without loss of generality since we can always repeat the following procedure for each connected component of $\mathcal{I}$ and $\mathcal{X} \setminus \mathcal{C}$.



\textbf{Initial Activation Set Identification: }
First, we identify the initial boundary activation set $\mathbf{S}_0$.
Given $\hat{x}_{U}\in \mathcal{X}\setminus\mathcal{C}$ and $\hat{x}_\mathcal{I}\in \mathcal{I}$, define a line segment
\begin{equation}
    \label{eq:lineseg}
    \tilde{L} = \left\{ x \in \mathbb{R}^n \mid x = (1 - \lambda) \hat{x}_{\mathcal{X}\setminus\mathcal{C}} + \lambda \hat{x}_\mathcal{I}, \; \lambda \in [0, 1] \right\}
\end{equation}
The following lemma shows the initial boundary activation set $\mathbf{S}_0$ can always be produced as $\mathbf{S}_0=\mathbf{S}(\tilde{x})$ for some $\tilde{x}\in \tilde{L}$. 

\begin{lemma}
\label{lemma:enu_init}
    Given two sample points $\hat{x}_U$, such that $b(\hat{x}_U)<0$, and $\hat{x}_\mathcal{I}$, such that $b(\hat{x}_U)>0$, let $\tilde{L}$ denote the line segment connecting these two points. Then, there exists a point $\tilde{x}\in \tilde{L}$ with $b_{\theta}(\tilde{x})=0$. 
\end{lemma}
Lemma 1 follows from the intermediate value theorem and continuity of $b_{\theta}(x)$. 
In order to search for $\mathbf{S}_0$, we choose a sequence of $N_{\tilde{L}}$ points $x_{1}^{0}, \ldots, x_{N_{\tilde{L}}}^{0} \in \tilde{L}$. For each $x_{i}^{0}$, we check to see if $\overline{\mathcal{X}}(\mathbf{S}(x_{i}^{0}))\cap \partial \mathcal{D}\neq \emptyset$ by solving boundary linear program $\text{BoundaryLP}(\mathbf{S}(x_{i}^{0}))$ \eqref{eq:boundarylp} in Appendix \ref{appendix:nlp_enum}

\textbf{Activation Sets Enumeration: }
We next describe Neural Breadth-First Search (NBFS) for enumerating all activation sets along the zero-level set, given an initial set $\mathbf{S}_0$. NBFS enumerates a collection of boundary hyperplanes denoted as $\mathcal{S}$ by repeating the following two steps. 

\textit{Step 1:} Given a set $\mathbf{S}$, NBFS identifies a collection of neighbor activation sets denoted as $\tilde{\mathcal{B}}$ as follows.  For each $(i,j)$ with $i=1,\ldots,L$ and $j=1,\ldots,M_{i}$, construct $\mathbf{S}^{\prime}$ as 
$\mathbf{S}^{\prime} = \begin{cases}\mathbf{S} \setminus {(i,j)}, & (i,j)\in \mathbf{S}^{\prime} \\ \mathbf{S} \cup {(i,j)}, & (i,j)\notin \mathbf{S}^{\prime} \end{cases}$. We check whether $\mathbf{S}^{\prime}\in \mathcal{B}$ by solving the  linear program $\text{USLP}(\mathbf{S}, (i,j))$ \eqref{eq:uslp} in Appendix \ref{appendix:nlp_enum}. 
 
If there exists such a state $x$, then $\mathbf{S}^{\prime}$ is added to $\tilde{\mathcal{B}}$. To further improve efficiency, we employ the simplex algorithm to calculate the hypercube that overapproximates the boundary hyperplane, denoted as $\mathcal{H}(\mathbf{S})\supseteq \overline{\mathcal{X}}(\mathbf{S})$, and relax the last constraint of \eqref{eq:boundarylp} to $x \in \mathcal{H}(\mathbf{S})$.

\textit{Step 2:} For each $\mathbf{S}^{\prime} \in \tilde{\mathcal{B}}$, NBFS  determines if the activation region $\overline{\mathcal{X}}(\mathbf{S}^{\prime})$ is on the boundary (i.e., satisfies $\overline{\mathcal{X}}(\mathbf{S}^{\prime}) \cap \partial \mathcal{D} \neq \emptyset$) by checking the feasibility of $BoundaryLP(\mathbf{S}^{\prime})$. If there exists such a state $x$, then $\mathbf{S}^{\prime}$ is added to $\mathcal{S}$. 
This process continues in a breadth-first search manner until all such activation sets on the boundary have been enumerated. When this phase terminates, $\mathcal{S} = \mathcal{B}$. Detailed procedure is described as Algorithm \ref{alg:NBFS} in Appendix \ref{appendix:algorithm_veri}. 

\textbf{Hinge Enumeration: } The verifier of \sysname enumerates a collection $\mathcal{V}$ of boundary hinges, where each boundary hinge $\mathbf{V} \in \mathcal{V}$ is a subset $\mathbf{S}_{1},\ldots,\mathbf{S}_{r}$ of $\mathcal{B}$ with $\overline{\mathcal{X}}(\mathbf{S}_{1}) \cap \overline{\mathcal{X}}(\mathbf{S}_{r}) \cap \partial \mathcal{D} \neq \emptyset$.


Given $\mathcal{S}_i \subseteq\mathcal{B}$ and $\mathbf{S}$, hinge enumeration filter the set of neighbor activation sets of $\mathbf{S}$ defined as $\mathcal{N}_{\mathbf{S}}(\mathcal{S}_i):=\{\mathbf{S}^{\prime}:\mathbf{S}^{\prime}\Delta \mathbf{S}= 1, \mathbf{S}^{\prime}\in \mathcal{S}_i\}$. Then, hinge enumeration identifies hinges $\mathbf{V}$ by solving linear program $\text{HingeLP}(\mathcal{N}^{(d)}_{\mathbf{S}}(\mathcal{S}_i))$ \eqref{eq:hingelp} in Appendix \ref{appendix:nlp_enum}. 
If $\exists x$, hinge enumeration includes the hinge into the set $\mathcal{V}\cup \mathbf{V}$. 
The efficiency can be further improved by leveraging the sufficient condition verification proposed in Section \ref{subsec:efficient_veri}. 
The following result describes the completeness guarantees of $\mathcal{S}$ and $\mathcal{V}$ enumerated in Line 5 and 8 of Algorithm \ref{alg:EEV}.

\begin{proposition}
\label{prop:enumeration-complete}
Let $\mathcal{S}$ and $\mathcal{V}$ denote the output of Algorithm \ref{alg:EEV}. Then the  boundary $\partial \mathcal{D}$ satisfies $\partial \mathcal{D} \subseteq \bigcup_{\mathbf{S} \in \mathcal{S}}{\overline{\mathcal{X}}(\mathbf{S})}$. Furthermore, if $\mathcal{S}$ is complete and $\left(\bigcap_{i=1}^{r}{\overline{\mathcal{X}}(\mathbf{S}_{i})}\right) \cap \{x: b(x) = 0\} \neq \emptyset$, then $\{\mathbf{S}_{1},\ldots,\mathbf{S}_{r}\} \in \mathcal{V}$.
\end{proposition}
The proof is omitted due to the space limit. A detailed proof is provided in Appendix \ref{appendix:proof_lemma_completeEnu}

\subsection{Efficient Verification}
\label{subsec:efficient_veri}


The efficient verification component takes the sets of boundary hyperplanes $\mathcal{S}$ and boundary hinges $\mathcal{V}$ and checks that the conditions of Proposition \ref{prop:safety-condition} hold for each of them. As pointed out after Proposition \ref{prop:safety-condition}, the problem of searching for safety counterexamples can be decomposed into searching for correctness, hyperplane, and hinge counterexamples. In order to maximize the efficiency of our approach, we first consider the least computationally burdensome verification task, namely, searching for correctness counterexamples. We then search for hyperplane counterexamples, followed by hinge counterexamples.


\textbf{Correctness Verification: }
The correctness condition (\eqref{eq:neural-containment}) can be verified for boundary hyperplane $\overline{\mathcal{X}}(\mathbf{S})$ by solving the  nonlinear program \eqref{eq:containment-verification} in Appendix \ref{appendix:nlp}. 
When $h(x)$ is convex, \eqref{eq:containment-verification} can be solved efficiently. Otherwise, dReal can be used to check satisfiability of \eqref{eq:containment-verification} in a tractable runtime.


\textbf{Hyperplane Verification: } 
Hyperplane counterexamples can be identified by solving the optimization problem \eqref{eq:hyperplane-verification-exact} in Appendix \ref{appendix:nlp}. 
Solving \eqref{eq:hyperplane-verification-exact} can be made more efficient when additional structures on the input set $\mathcal{U}$ and dynamics $f$ and $g$ are present. Consider a case $\mathcal{U} = \{D\omega : ||\omega||_{\infty} \leq 1\}$. In this case, the problem reduces to the  nonlinear program \eqref{eq:interval_minLf} in Appendix \ref{appendix:nlp}. 
If $f(x)$ and $g(x)$ are linear in $x$, then the problem boils down to a linear program. If bounds on the Lipschitz coefficients of $f$ and $g$ are available, then they can be used to derive approximations of \eqref{eq:interval_minLf}.

If $\mathcal{U} = \mathbb{R}^{m}$, then  by \cite[Corollary 1]{zhang2023exact}, the problem can be reduced to the  nonlinear program \eqref{eq:single-set-nonlinear-prog-special-case} in Appendix \ref{appendix:nlp}. 
If $g(x)$ is a constant matrix $G$, then safety is automatically guaranteed if $\overline{W}(\mathbf{S})^{T}G \neq 0$. If $f(x)$ is linear in $x$ as well, then (\ref{eq:single-set-nonlinear-prog-special-case}) is a linear program. 

\textbf{Hinge Verification: }
The hinge $\mathbf{V} = \{\mathbf{S}_{1},\ldots,\mathbf{S}_{r}\}$ can be certified by solving the nonlinear optimization problem \eqref{eq:hinge-certify} in Appendix \ref{appendix:nlp}. 
In practice, simple heuristics are often sufficient to verify safety of hinges without resorting to solving \eqref{eq:hinge-certify}. If $\overline{W}(\mathbf{S})^{T}f(x) > 0$ for all $x \in \overline{\mathcal{X}}(\mathbf{S}_{1}) \cap \cdots \cap \overline{\mathcal{X}}(\mathbf{S}_{r})$, then the control input $u = \mathbf{0}$ suffices to ensure safety. Furthermore, if $\mathcal{U} = \mathbb{R}^{m}$ and there exists $i \in \{1,\ldots,m\}$ and $s \in \{0,1\}$ such that $\mbox{sign}((\overline{W}(\mathbf{S}_{l})^{T}g(x))_{i}) = s$ for all $x \in \overline{\mathcal{X}}(\mathbf{S}_{1}) \cap \cdots \overline{\mathcal{X}}(\mathbf{S}_{r})$ and $l=1,\ldots,r$, then $u$ can be chosen as $u_{i} = Ks$ for some sufficiently large $K > 0$ to ensure that the conditions of Proposition \ref{prop:safety-condition}.



\textbf{Safety Guarantee: } The safety guarantees of our proposed approach are summarized in Theorem \ref{th:verification_guarantee}.

\begin{theorem}
\label{th:verification_guarantee}
Given a NCBF $b_{\theta}(x)$, $b_{\theta}(x)$ is a valid NCBF if it passes the verification of  Algorithm \ref{alg:EEV} using dReal to solve the optimization problems \eqref{eq:containment-verification}, \eqref{eq:hyperplane-verification-exact}, and \eqref{eq:hinge-certify}.
\end{theorem}
The proof is derived from completeness of the enumeration in Algorithm \ref{alg:EEV} and dReal. A detailed proof can be found in Appendix \ref{appendix:veri_guarantee}

\section{Experiments}
\label{sec:experiments}

In this section, we evaluate the proposed \sysname in regularizer efficacy and verification efficiency. We also demonstrated the improved performance with counter-example guidance in the synthesis framework, whose results are detailed in \ref{appendix:synthesis_framework}. We experiment on four systems, namely Darboux, obstacle avoidance, hi-ord$_8$, and spacecraft rendezvous. The experiments run on a workstation with an Intel i7-11700KF CPU, and an NVIDIA GeForce RTX 3080 GPU. We include experiment settings in Appendix \ref{appendix:exp_settings} and hyperparameters settings in Table \ref{tab:hyperparameter} in Appendix \ref{appendix:hyper_para}.

\subsection{Experiment Setup}
\textbf{Darboux:} We consider the Darboux system~\cite{zeng2016darboux}, a nonlinear open-loop polynomial system with detailed settings presented in Appendix \ref{appendix:exp_settings} 

\textbf{Obstacle Avoidance (OA):} We evaluate our proposed method on a controlled system~\cite{barry2012safety}. We consider Unmanned Aerial Vehicles (UAVs) avoiding collision with a tree trunk. The system state consists of a 2-D position and aircraft yaw rate $x:=[x_1, x_2, \psi]^T$. The system is manipulated by the yaw rate input $u$ with detailed settings presented in Appendix \ref{appendix:exp_settings}. 

\textbf{Spacecraft Rendezvous (SR):} We evaluate our approach on a spacecraft rendezvous problem from~\cite{jewison2016spacecraft}. The system state is $x=[p_x, p_y, p_z, v_x, v_y, v_z]^T$ and control input is $u=[u_x, u_y, u_z]^T$ with with detailed settings presented in Appendix \ref{appendix:exp_settings}.  

\textbf{hi-ord$_8$:} We evaluate our approach on an eight-dimensional system that first appeared in \cite{abate2021fossil} to evaluate the scalability of the proposed method. Detailed settings can be found in Appendix \ref{appendix:exp_settings}

\subsection{Regularizer Efficacy Evaluation}


\begin{figure}[t]
    \centering
    
    \begin{subfigure}[b]{0.35\textwidth}
        \centering
        \includegraphics[trim={0 0.1in 0 0.35in},clip,width=\textwidth]{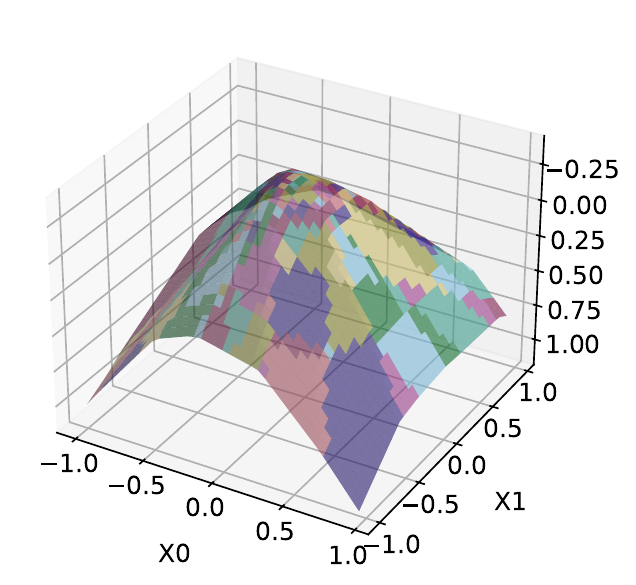}
        \caption{$r=0$}
        \label{fig:reg0}
    \end{subfigure}
    \begin{subfigure}[b]{0.35\textwidth}
        \centering
        \includegraphics[trim={0 0.1in 0 0.35in},clip,width=\textwidth]{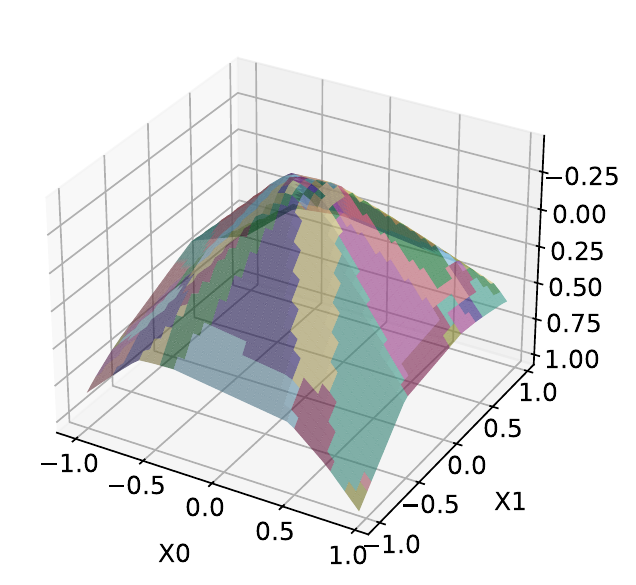}
        \caption{$r=50$}
        \label{fig:reg50}
    \end{subfigure}
    \caption{Effects of boundary regularization ($r$) on activation sets along the boundary. The figures show the results from a neural network with 4 layers of 8 hidden units, applied to the Spacecraft case. The surface represents the first two dimensions with the last four dimensions fixed at 0. Increasing $r$ results in more organized boundary activation sets.}
    \label{fig:boundary_regularization}
    \vspace{-0.2in}
\end{figure}

\begin{table}[!htbp]
\caption{Comparison of $N$ the number of boundary hyperplanes and $C$ coverage of the safe region $\mathcal{D}$ of NCBF trained with and without boundary hyperplane regularizer denoted with subscripts $_r$ and $_o$.}  
\label{tab:NCBFcomparison}
\centering
\begin{tabular}{llllllllllll}
\toprule
Case  & n & L & M  & $N_o$ & $C_o$ & $N_{r=1}$ & $\rho_{r=1}$ & $N_{r=10}$ & $\rho_{r=10}$ & $N_{r=50}$ & $\rho_{r=50}$ \\ 
\cmidrule(r){1-12}
 \multirow{ 4}{*}{OA} & 3 & 2 & 8 & 26 & 89.46\% & 25 & 0.996 & 23.3 & 0.994 & \textbf{13.3} & 1.006  \\
 & 3 & 2 & 16 & 116 & 83.74\% & 119 & 1.012 & 111 & 1.005 & \textbf{98} & 1.055  \\
 & 3 & 4 & 8  & 40 & 91.94\% & 38 & 0.988 & 36 & 0.993 & \textbf{13} & 0.937  \\
 & 3 & 4 & 16 & 156 & 87.81\% & 170 & 0.971 & 147 & 1.003 & \textbf{64} & 1.038  \\
\cmidrule(r){1-12}
\multirow{ 3}{*}{SR}
& 6 & 2 & 8 & 2868 & 98.58\% & 2753  & 1 & 1559 & 1 & \textbf{418} & 1  \\ 
& 6 & 4 & 8 & 6371 & 98.64\% & 6218  & 1 & 3055 & 1 & \textbf{627} & 1  \\ 
& 6 & 2 & 16 & N/A & N/A & 204175 & N/A & 68783 & N/A & \textbf{13930} & N/A  \\ 
\bottomrule
\end{tabular}
\end{table}


Table~\ref{tab:NCBFcomparison} and Figure~\ref{fig:boundary_regularization} illustrates the impact of regularization on the CBF boundary's activation sets. Table~\ref{tab:NCBFcomparison} compares various configurations, where \( n \) denotes the input dimensions, \( L \) represents the number of layers, and \( M \) indicates the number of hidden units per layer. \( N_o \) and \( N_r \) are the number of hyperplanes along the zero-level boundary of the CBF without and with regularization, respectively, with \( r \) indicating the regularization strength. \( C_o \) captures the CBF's coverage of the safe region, while \( \rho_{(\cdot)} = C_{(\cdot)}/C_o \) represents the safety coverage ratio relative to the unregularized CBF. Notably, "N/A" entries indicate configurations where training a fully verifiable network was infeasible due to the excessive number of boundary hyperplanes, which leads the verification process to time out.

The results demonstrate that regularization effectively reduces the number of activation sets along the CBF boundary without compromising the coverage of the safe region. The efficiency is especially improved in cases with a greater number of hidden layers, where the unregularized model results in a significantly higher number of hyperplanes. For instance, in the SR case with \( n=6 \), \( L=4 \), and \( M=8 \), the regularization reduces \( N_{r=50} \) to $627$ from \( N_o=6218 \), maintaining the same safety coverage ($\rho_{r=50} = 1$). See Appendix \ref{appendix:sensitivity} for hyperparameter sensitivity analysis.

Figure~\ref{fig:boundary_regularization} illustrates the level sets of two CBFs trained for the SR case with \( n=6 \), \( L=4 \), and \( M=8 \). These level sets are extracted from the first two dimensions with the rest set to zero. Each colored patch represents an activation pattern. The regularizer organizes the activation sets around the boundary, reducing unnecessary rapid changes and thereby enhancing the 
verification efficiency. 

\subsection{Efficient Verification Evaluation}


\begin{table}[!htbp]
\caption{Comparison of verification run-time of NCBF in seconds. We denote the run-time as `UTD' when the method is unable to be directly used for verification. } 
\label{tab:NCBFcomparison}
\centering
\begin{tabular}{lllllllllll}
\toprule
Case  & n & L & M  & $N$ & $t_h$ & $t_g$  & \sysname & Baseline~\cite{zhang2023exact} & dReal & Z3 \\ \cmidrule(r){1-11}
\multirow{ 2}{*}{Darboux} & 2 & 2 & 256  & 15 & 2.5s & 0 & 2.5s & 315s & >3h & >3h \\
& 2 & 2 & 512  & 15 & 3.3s & 0 & 3.3s & 631s & >3h & >3h \\
\cmidrule(r){1-11}
\multirow{ 4}{*}{OA} & 3 & 2 & 16 & 86 & 0.41s & 0 & 0.41s & 16.0s & >3h & >3h \\
 & 3 & 4 & 8 & 15 & 0.39 & 0 & 0.39 & 16.1s & >3h & >3h \\
 & 3 & 4 & 16 & 136 & 0.65s & 0 & 0.65s & 36.7s & >3h & >3h \\
 & 3 & 1 & 128 & 5778 & 20.6s & 0 & 20.6s & 207s & >3h & >3h \\
\cmidrule(r){1-11}
\multirow{ 3}{*}{hi-ord$_8$} & 8 & 2 & 8 & 73 & 0.54s & 0 & 0.54s & >3h & >3h & >3h \\
 & 8 & 2 & 16 & 3048 & 11.8s & 0 & 11.8s & >3h & >3h & >3h \\
 & 8 & 4 & 16 & 3984 & 22.4s & 0 & 22.4s & >3h & >3h & >3h \\
\cmidrule(r){1-11}
\multirow{ 2}{*}{SR}
& 6 & 2 & 8 & 2200 & 7.1s & 2.7s  & 9.8s & 179s & UTD & UTD \\ 
& 6 & 4 & 8 & 4918 & 45.8s & 14.3s  & 60.1s & 298.7s & UTD & UTD \\ 
\bottomrule
\end{tabular}
\label{tab:verification_efficiency}
\end{table}

The results presented in Table~\ref{tab:verification_efficiency} illustrate a significant improvement in verification efficiency for Neural Control Barrier Functions NCBFs using the proposed method. In the table \( t_h \) represents the time spent searching for hyperplanes containing the CBF boundary and verifying CBF sufficient conditions on these boundaries, and \( t_g \) represents the time spent in hinge enumeration and verification. The total time, \( T \), is the sum of \( t_h \) and \( t_g \). We compare our approach with three baselines, exact verification \cite{zhang2023exact}, SMT-based verification \cite{abate2021fossil} with dReal and Z3.  Baseline methods' run times are represented by Baseline \cite{zhang2023exact}, dReal, and Z3.

In the Darboux cases, our method achieves verification in 2.5 seconds and 3.3 seconds for \( M = 256 \) and \( M = 512 \) respectively, whereas baseline methods take substantially longer, with Baseline~\cite{zhang2023exact} taking 315 seconds and 631 seconds, and both dReal and Z3 taking more than 3 hours. Similarly, in the OA cases, our method's run times range from 0.39 seconds to 20.6 seconds, faster than the baseline methods. 
In the more higher dimensional systems high-ord$_8$ and SR, our method significantly outperforms Baseline~\cite{zhang2023exact}. Specifically, in high-ord$_8$ our methods finishes within 22.4 seconds while Baseline~\cite{zhang2023exact}, dReal and Z3 times out, due to the need to enumerate the 8-dimensional input space. For the SR case, SEEV's run time are 9.8 seconds and 60.1 seconds, beating Baseline~\cite{zhang2023exact} which takes 179 seconds and 298.7 seconds respectively. Neural barrier certificate based dReal and Z3 are able to directly applicable since they require an explicit expression of the controlled feedback system. However, the SR system is manipulated by an NCBF-based safe controller that is nontrivial to derive an explicit expression. 

Note that hinge enumeration and certification may be time-consuming procedure, since they involve enumerating all combinations of hyperplanes. However, the results from Table~\ref{tab:verification_efficiency} show that the certification can be completed on most hyperplanes with sufficient condition verification in Section \ref{subsec:efficient_veri}, greatly improving the overall run time.


\section{Conclusion}
\label{sec:conclusions}
This paper considered the problem of synthesizing and verifying NCBFs with ReLU activation function in an efficient manner. Our approach is guided by the fact that the main contribution to the computational cost of verifying NCBFs is enumerating and verifying safety of each piecewise-linear activation region at the safety boundary. We proposed Synthesis with Efficient Exact Verification (SEEV), which co-designs the synthesis and verification components to enhance scalability of the verification process. We augment the NCBF synthesis with a regularizer that reduces the number of piecewise-linear segments at the boundary, and hence reduces the total workload of the verification. We then propose a verification approach that efficiently enumerates the linear segments at the boundary and exploits tractable sufficient conditions for safety.  

\textbf{Limitations:} 
The method proposed in this paper mitigated the scalability issue. However, the synthesis and verification of NCBFs for higher-dimensional systems is challenging. Exact verification of non-ReLU NCBFs, which lack ReLU's simple piecewise linearity, remains an open problem. 

\section*{Acknowledgements and Disclosure of Funding}
This research was partially supported by NSF grant CNS-1941670, CMMI-2418806, AFOSR grant FA9550-22-1-0054, NSF Career CCF 2047034, NSF CCF DASS 2217723, NSF AI Institute CCF 2112665, and Amazon Research Award.

\bibliographystyle{unsrt}
\bibliography{ref}

\begin{thebibliography}{10}

\bibitem{hsu2015control}
Shao-Chen Hsu, Xiangru Xu, and Aaron~D Ames.
\newblock Control barrier function based quadratic programs with application to
  bipedal robotic walking.
\newblock In {\em 2015 American Control Conference (ACC)}, pages 4542--4548.
  IEEE, 2015.

\bibitem{agrawal2021safe}
Devansh~R Agrawal and Dimitra Panagou.
\newblock Safe control synthesis via input constrained control barrier
  functions.
\newblock In {\em 2021 60th IEEE Conference on Decision and Control (CDC)},
  pages 6113--6118. IEEE, 2021.

\bibitem{xu2017correctness}
Xiangru Xu, Jessy~W Grizzle, Paulo Tabuada, and Aaron~D Ames.
\newblock Correctness guarantees for the composition of lane keeping and
  adaptive cruise control.
\newblock {\em IEEE Transactions on Automation Science and Engineering},
  15(3):1216--1229, 2017.

\bibitem{breeden2021high}
Joseph Breeden and Dimitra Panagou.
\newblock High relative degree control barrier functions under input
  constraints.
\newblock In {\em 2021 60th IEEE Conference on Decision and Control (CDC)},
  pages 6119--6124. IEEE, 2021.

\bibitem{dai2023convex}
Hongkai Dai and Frank Permenter.
\newblock Convex synthesis and verification of control-lyapunov and barrier
  functions with input constraints.
\newblock In {\em 2023 American Control Conference (ACC)}, pages 4116--4123.
  IEEE, 2023.

\bibitem{kang2023verification}
Shucheng Kang, Yuxiao Chen, Heng Yang, and Marco Pavone.
\newblock Verification and synthesis of robust control barrier functions:
  Multilevel polynomial optimization and semidefinite relaxation, 2023.

\bibitem{rober2023hybrid}
Nicholas Rober, Michael Everett, Songan Zhang, and Jonathan~P How.
\newblock A hybrid partitioning strategy for backward reachability of neural
  feedback loops.
\newblock In {\em 2023 American Control Conference (ACC)}, pages 3523--3528.
  IEEE, 2023.

\bibitem{dawson2023safe}
Charles Dawson, Sicun Gao, and Chuchu Fan.
\newblock Safe control with learned certificates: A survey of neural
  {L}yapunov, barrier, and contraction methods for robotics and control.
\newblock {\em IEEE Transactions on Robotics}, 2023.

\bibitem{so2023train}
Oswin So, Zachary Serlin, Makai Mann, Jake Gonzales, Kwesi Rutledge, Nicholas
  Roy, and Chuchu Fan.
\newblock How to train your neural control barrier function: Learning safety
  filters for complex input-constrained systems.
\newblock {\em arXiv preprint arXiv:2310.15478}, 2023.

\bibitem{dawson2022safe}
Charles Dawson, Zengyi Qin, Sicun Gao, and Chuchu Fan.
\newblock Safe nonlinear control using robust neural {L}yapunov-barrier
  functions.
\newblock In {\em Conference on Robot Learning}, pages 1724--1735. PMLR, 2022.

\bibitem{tayal2024learning}
Manan Tayal, Hongchao Zhang, Pushpak Jagtap, Andrew Clark, and Shishir
  Kolathaya.
\newblock Learning a formally verified control barrier function in stochastic
  environment.
\newblock {\em arXiv preprint arXiv:2403.19332}, 2024.

\bibitem{long2021learning}
Kehan Long, Cheng Qian, Jorge Cort{\'e}s, and Nikolay Atanasov.
\newblock Learning barrier functions with memory for robust safe navigation.
\newblock {\em IEEE Robotics and Automation Letters}, 6(3):4931--4938, 2021.

\bibitem{xiao2023barriernet}
Wei Xiao, Tsun-Hsuan Wang, Ramin Hasani, Makram Chahine, Alexander Amini, Xiao
  Li, and Daniela Rus.
\newblock Barriernet: Differentiable control barrier functions for learning of
  safe robot control.
\newblock {\em IEEE Transactions on Robotics}, 2023.

\bibitem{zhao2021learning}
Hengjun Zhao, Xia Zeng, Taolue Chen, Zhiming Liu, and Jim Woodcock.
\newblock Learning safe neural network controllers with barrier certificates.
\newblock {\em Formal Aspects of Computing}, 33:437--455, 2021.

\bibitem{zhang2018efficient}
Huan Zhang, Tsui-Wei Weng, Pin-Yu Chen, Cho-Jui Hsieh, and Luca Daniel.
\newblock Efficient neural network robustness certification with general
  activation functions.
\newblock {\em Advances in Neural Information Processing Systems},
  31:4939--4948, 2018.

\bibitem{xu2020automatic}
Kaidi Xu, Zhouxing Shi, Huan Zhang, Yihan Wang, Kai-Wei Chang, Minlie Huang,
  Bhavya Kailkhura, Xue Lin, and Cho-Jui Hsieh.
\newblock Automatic perturbation analysis for scalable certified robustness and
  beyond.
\newblock {\em Advances in Neural Information Processing Systems}, 33, 2020.

\bibitem{salman2019convex}
Hadi Salman, Greg Yang, Huan Zhang, Cho-Jui Hsieh, and Pengchuan Zhang.
\newblock A convex relaxation barrier to tight robustness verification of
  neural networks.
\newblock {\em Advances in Neural Information Processing Systems},
  32:9835--9846, 2019.

\bibitem{xu2021fast}
Kaidi Xu, Huan Zhang, Shiqi Wang, Yihan Wang, Suman Jana, Xue Lin, and Cho-Jui
  Hsieh.
\newblock {Fast and Complete}: Enabling complete neural network verification
  with rapid and massively parallel incomplete verifiers.
\newblock In {\em International Conference on Learning Representations}, 2021.

\bibitem{wang2021beta}
Shiqi Wang, Huan Zhang, Kaidi Xu, Xue Lin, Suman Jana, Cho-Jui Hsieh, and
  J~Zico Kolter.
\newblock {Beta-CROWN}: Efficient bound propagation with per-neuron split
  constraints for complete and incomplete neural network verification.
\newblock {\em Advances in Neural Information Processing Systems}, 34, 2021.

\bibitem{zhang22babattack}
Huan Zhang, Shiqi Wang, Kaidi Xu, Yihan Wang, Suman Jana, Cho-Jui Hsieh, and
  Zico Kolter.
\newblock A branch and bound framework for stronger adversarial attacks of
  {R}e{LU} networks.
\newblock In {\em Proceedings of the 39th International Conference on Machine
  Learning}, volume 162, pages 26591--26604, 2022.

\bibitem{abate2021fossil}
Alessandro Abate, Daniele Ahmed, Alec Edwards, Mirco Giacobbe, and Andrea
  Peruffo.
\newblock Fossil: A software tool for the formal synthesis of {L}yapunov
  functions and barrier certificates using neural networks.
\newblock In {\em Proceedings of the 24th International Conference on Hybrid
  Systems: Computation and Control}, pages 1--11, 2021.

\bibitem{edwards2023fossil}
Alec Edwards, Andrea Peruffo, and Alessandro Abate.
\newblock Fossil 2.0: Formal certificate synthesis for the verification and
  control of dynamical models.
\newblock {\em arXiv preprint arXiv:2311.09793}, 2023.

\bibitem{zhang2023exact}
Hongchao Zhang, Junlin Wu, Yevgeniy Vorobeychik, and Andrew Clark.
\newblock Exact verification of relu neural control barrier functions.
\newblock {\em Advances in Neural Information Processing Systems}, 36, 2024.

\bibitem{berkenkamp2017safe}
Felix Berkenkamp, Matteo Turchetta, Angela Schoellig, and Andreas Krause.
\newblock Safe model-based reinforcement learning with stability guarantees.
\newblock {\em Advances in {N}eural {I}nformation {P}rocessing {S}ystems}, 30,
  2017.

\bibitem{qin2021learning}
Zengyi Qin, Kaiqing Zhang, Yuxiao Chen, Jingkai Chen, and Chuchu Fan.
\newblock Learning safe multi-agent control with decentralized neural barrier
  certificates.
\newblock {\em arXiv preprint arXiv:2101.05436}, 2021.

\bibitem{qin2022quantifying}
Zhizhen Qin, Tsui-Wei Weng, and Sicun Gao.
\newblock Quantifying safety of learning-based self-driving control using
  almost-barrier functions.
\newblock In {\em 2022 IEEE/RSJ International Conference on Intelligent Robots
  and Systems (IROS)}, pages 12903--12910. IEEE, 2022.

\bibitem{liu2023safe}
Simin Liu, Changliu Liu, and John Dolan.
\newblock Safe control under input limits with neural control barrier
  functions.
\newblock In {\em Conference on Robot Learning}, pages 1970--1980. PMLR, 2023.

\bibitem{prajna2007framework}
Stephen Prajna, Ali Jadbabaie, and George~J Pappas.
\newblock A framework for worst-case and stochastic safety verification using
  barrier certificates.
\newblock {\em IEEE Transactions on Automatic Control}, 52(8):1415--1428, 2007.

\bibitem{ames2019control}
Aaron~D Ames, Samuel Coogan, Magnus Egerstedt, Gennaro Notomista, Koushil
  Sreenath, and Paulo Tabuada.
\newblock Control barrier functions: Theory and applications.
\newblock In {\em 2019 18th European control conference (ECC)}, pages
  3420--3431. IEEE, 2019.

\bibitem{zhao2023safety}
Weiye Zhao, Tairan He, Tianhao Wei, Simin Liu, and Changliu Liu.
\newblock Safety index synthesis via sum-of-squares programming.
\newblock In {\em 2023 American Control Conference (ACC)}, pages 732--737.
  IEEE, 2023.

\bibitem{schneeberger2023sos}
Michael Schneeberger, Florian D{\"o}rfler, and Silvia Mastellone.
\newblock {SOS} construction of compatible control {L}yapunov and barrier
  functions.
\newblock {\em arXiv preprint arXiv:2305.01222}, 2023.

\bibitem{clark2021verification}
Andrew Clark.
\newblock Verification and synthesis of control barrier functions.
\newblock In {\em 2021 60th IEEE Conference on Decision and Control (CDC)},
  pages 6105--6112. IEEE, 2021.

\bibitem{ferrari2022complete}
Claudio Ferrari, Mark~Niklas Muller, Nikola Jovanovic, and Martin Vechev.
\newblock Complete verification via multi-neuron relaxation guided
  branch-and-bound.
\newblock {\em arXiv preprint arXiv:2205.00263}, 2022.

\bibitem{henriksen2021deepsplit}
Patrick Henriksen and Alessio Lomuscio.
\newblock Deepsplit: An efficient splitting method for neural network
  verification via indirect effect analysis.
\newblock In {\em IJCAI}, pages 2549--2555, 2021.

\bibitem{zhang2022general}
Huan Zhang, Shiqi Wang, Kaidi Xu, Linyi Li, Bo~Li, Suman Jana, Cho-Jui Hsieh,
  and J~Zico Kolter.
\newblock General cutting planes for bound-propagation-based neural network
  verification.
\newblock {\em Advances in Neural Information Processing Systems},
  35:1656--1670, 2022.

\bibitem{katz2017reluplex}
Guy Katz, Clark Barrett, David~L Dill, Kyle Julian, and Mykel~J Kochenderfer.
\newblock Reluplex: An efficient {SMT} solver for verifying deep neural
  networks.
\newblock In {\em Computer Aided Verification: 29th International Conference,
  CAV 2017, Heidelberg, Germany, July 24-28, 2017, Proceedings, Part I 30},
  pages 97--117. Springer, 2017.

\bibitem{katz2019marabou}
Guy Katz, Derek~A Huang, Duligur Ibeling, Kyle Julian, Christopher Lazarus,
  Rachel Lim, Parth Shah, Shantanu Thakoor, Haoze Wu, Aleksandar Zelji{\'c},
  et~al.
\newblock The {M}arabou framework for verification and analysis of deep neural
  networks.
\newblock In {\em Computer Aided Verification: 31st International Conference,
  CAV 2019, New York City, NY, USA, July 15-18, 2019, Proceedings, Part I 31},
  pages 443--452. Springer, 2019.

\bibitem{zhao2020synthesizing}
Hengjun Zhao, Xia Zeng, Taolue Chen, and Zhiming Liu.
\newblock Synthesizing barrier certificates using neural networks.
\newblock In {\em Proceedings of the 23rd international conference on hybrid
  systems: Computation and control}, pages 1--11, 2020.

\bibitem{abate2020formal}
Alessandro Abate, Daniele Ahmed, Mirco Giacobbe, and Andrea Peruffo.
\newblock Formal synthesis of lyapunov neural networks.
\newblock {\em IEEE Control Systems Letters}, 5(3):773--778, 2020.

\bibitem{10.1007/978-3-030-72016-2_20}
Andrea Peruffo, Daniele Ahmed, and Alessandro Abate.
\newblock Automated and formal synthesis of neural barrier certificates for
  dynamical models.
\newblock In {\em Tools and Algorithms for the Construction and Analysis of
  Systems}, pages 370--388. Springer International Publishing, 2021.

\bibitem{anand2023formally}
Mahathi Anand and Majid Zamani.
\newblock Formally verified neural network control barrier certificates for
  unknown systems.
\newblock {\em IFAC-PapersOnLine}, 56(2):2431--2436, 2023.

\bibitem{wang2022design}
Yixuan Wang, Chao Huang, Zhaoran Wang, Zhilu Wang, and Qi~Zhu.
\newblock Design-while-verify: correct-by-construction control learning with
  verification in the loop.
\newblock In {\em Proceedings of the 59th ACM/IEEE Design Automation
  Conference}, pages 925--930, 2022.

\bibitem{wang2024simultaneous}
Xinyu Wang, Luzia Knoedler, Frederik~Baymler Mathiesen, and Javier Alonso-Mora.
\newblock Simultaneous synthesis and verification of neural control barrier
  functions through branch-and-bound verification-in-the-loop training.
\newblock In {\em 2024 European Control Conference (ECC)}, pages 571--578.
  IEEE, 2024.

\bibitem{blanchini2008set}
Franco Blanchini and Stefano Miani.
\newblock {\em Set-{T}heoretic {M}ethods in {C}ontrol}, volume~78.
\newblock Springer, 2008.

\bibitem{cvxpylayers2019}
A.~Agrawal, B.~Amos, S.~Barratt, S.~Boyd, S.~Diamond, and Z.~Kolter.
\newblock Differentiable convex optimization layers.
\newblock In {\em Advances in Neural Information Processing Systems}, 2019.

\bibitem{zeng2016darboux}
Xia Zeng, Wang Lin, Zhengfeng Yang, Xin Chen, and Lilei Wang.
\newblock Darboux-type barrier certificates for safety verification of
  nonlinear hybrid systems.
\newblock In {\em Proceedings of the 13th International Conference on Embedded
  Software}, pages 1--10, 2016.

\bibitem{barry2012safety}
Andrew~J Barry, Anirudha Majumdar, and Russ Tedrake.
\newblock Safety verification of reactive controllers for uav flight in
  cluttered environments using barrier certificates.
\newblock In {\em 2012 IEEE International Conference on Robotics and
  Automation}, pages 484--490. IEEE, 2012.

\bibitem{jewison2016spacecraft}
Christopher Jewison and R~Scott Erwin.
\newblock A spacecraft benchmark problem for hybrid control and estimation.
\newblock In {\em 2016 IEEE 55th Conference on Decision and Control (CDC)},
  pages 3300--3305. IEEE, 2016.

\end{thebibliography}
\newpage
\appendix
\section{Supplement}
\subsection{Definition of $\overline{W}_{ij}(\mathbf{S})$ and $\overline{r}_{ij}(\mathbf{S})$}
\label{appendix:prelim}
In what follows, we define vectors $\overline{W}_{ij}(\mathbf{S}) \in \mathbb{R}^{n}$ and scalars $\overline{r}_{ij}(\mathbf{S}) \in \mathbb{R}$ such that $z_{j}^{(i)} = \overline{W}_{ij}(\mathbf{S})^{T}x + \overline{r}_{ij}(\mathbf{S})$. The weight and bias of the input layer is defined by 
\begin{displaymath}
    \overline{W}_{1j}(\mathbf{S}) = \left\{
    \begin{array}{ll}
    W_{1j}, & (1,j) \in \mathbf{S} \\
    0, & \mbox{else}
    \end{array}
    \right. \quad 
    \overline{r}_{1j}(\mathbf{S}) = \left\{
    \begin{array}{ll}
    r_{1j}, & (1,j) \in \mathbf{S} \\
    0, & \mbox{else}
    \end{array}
    \right.
\end{displaymath}
Proceeding inductively, define $\overline{W}_{i}(\mathbf{S}) \in \mathbb{R}^{M_{i} \times n}$ to be a matrix with columns $\overline{W}_{ij}(\mathbf{S})$ for $j=1,\ldots,M_{i}$ and $\overline{r}_{i}(\mathbf{S}) \in \mathbb{R}^{M_{i}}$ to be a vector with elements $\overline{r}_{ij}(\mathbf{S})$.  We then define 
\begin{displaymath}
\overline{W}_{ij}(\mathbf{S}) = \left\{
\begin{array}{ll}
\overline{W}_{i-1}(\mathbf{S})W_{ij}, & (i,j) \in \mathbf{S} \\
0, & \mbox{else}
\end{array}
\right. \quad
\overline{r}_{ij}(\mathbf{S}) = \left\{
\begin{array}{ll}
W_{ij}^{T}\overline{r}_{i-1}(\mathbf{S}) + r_{ij}, & (i,j) \in \mathbf{S} \\
0, & \mbox{else}
\end{array}
\right.
\end{displaymath}
At the last layer, let $\overline{W}(\mathbf{S}) = W_{L}(\mathbf{S})\Omega$ and $\overline{r}(\mathbf{S}) = \Omega^{T}r_{L}(\mathbf{S}) + \psi$, so that $y = \overline{W}(\mathbf{S})^{T}x + \overline{r}(\mathbf{S})$ if $\mathbf{S} = \{(i,j): z_{j}^{(i)} \geq 0\}$. 



\subsection{Proof of Proposition \ref{prop:enumeration-complete}}
\label{appendix:proof_lemma_completeEnu}
\begin{proof}
Suppose that $x^{\prime} \in \partial \mathcal{D} \setminus \left(\bigcup_{\mathbf{S} \in \mathcal{S}}{\overline{\mathcal{X}}(\mathbf{S})}\right)$, and let $x$ denote the state on $\partial \mathcal{D}$ found by Line 5 of Algorithm \ref{alg:EEV}. Since $\partial \mathcal{D}$ is connected, there exists a path $\gamma$ with $\gamma(0) = x$ and $\gamma(1)=x^{\prime}$ contained in $\partial \mathcal{D}$. Let $\mathbf{S}_{0}, \mathbf{S}_{1},\ldots,\mathbf{S}_{K}$ denote a sequence of activation sets with $\mathbf{S}_{0} \in \mathbf{S}(x)$, $\mathbf{S}_{K} \in \mathbf{S}(x^{\prime})$, and $\mathbf{S}_{0}$ equal to the set computed at Line 5 of Algorithm \ref{alg:EEV}. Then there exists $i \geq 2$ such that $\mathbf{S}_{i-1} \in \mathcal{S}$ and $\mathbf{S}_{i} \notin \mathcal{S}$, and there exists $t^{\prime}$ such that $\gamma(t^{\prime}) \in \overline{\mathcal{X}}(\mathbf{S}_{i-1}) \cap \overline{\mathcal{X}}(\mathbf{S}_{i}) \cap \{x: b(x) = 0\}$. We then have that $\mathbf{T}(\mathbf{S}_{i-1},\mathbf{S}_{i})$ is a subset of the set $\mathbf{T}$, and hence $\mathbf{S}_{i}$ will be identified and added to $\mathcal{S}$ at Line 5 of Algorithm \ref{alg:EEV}. 

Now, suppose that $\mathbf{S}_{1},\ldots,\mathbf{S}_{r} \in \mathcal{S}$ is complete and $\left(\bigcap_{i=1}^{r}{\overline{\mathcal{X}}_{0}(\mathbf{S}_{i})}\right) \cap \{x: b(x) = 0\} \neq \emptyset$. Let $\mathbf{T} = \mathbf{T}(\mathbf{S}_{1},\ldots,\mathbf{S}_{r})$. Then $\mathbf{T}$ is a subset of the sets $\mathbf{S}_{1},\ldots,\mathbf{S}_{r} \in \mathcal{S}$. Since the intersection of the $\overline{\mathcal{X}}(\mathbf{S})$ sets with $\{x: b(x) = 0\}$ is nonempty, $\{\mathbf{S}_{1},\ldots,\mathbf{S}_{r}\}$ is added to $\mathcal{V}$. 
\end{proof}

\subsection{Algorithm for Verification}
\label{appendix:algorithm_veri}

\sysname identifies the initial activation set by conducting a BoundaryLP-based binary search as shown in Algorithm. \ref{alg:identify}. 
The algorithm presents a procedure to identify a hyperplane characterized by an initial activation set $\mathbf{S}_0$ that may contain the boundary $\partial \mathcal{D}$. 
It iterates over pairs of sample states from the unsafe training set $\mathcal{T}_{\mathcal{X} \setminus \mathcal{C}}$ and the safe training set $\mathcal{T}_{\mathcal{I}}$. For each pair $(\hat{x}_{\mathcal{X} \setminus \mathcal{C}}, \hat{x}_{\mathcal{I}})$, the algorithm initializes the left and right points of a search interval. 
It then performs a binary search by repeatedly computing the midpoint $x_{\text{mid}}$ and checking the feasibility of the boundary linear program $\text{BoundaryLP}(x_{\text{mid}})$. 
The algorithm terminates when $\text{BoundaryLP}(x_{\text{mid}})$ is feasible, returning the activation set $\mathbf{S}(x_{\text{mid}})$ as $\mathbf{S}_0$.

\begin{algorithm}[h]
\caption{Binary Search for $\mathbf{S}_{0}$}
\begin{algorithmic}[1]
    \State \textbf{Input:} $\mathcal{T}_{\mathcal{X}\setminus\mathcal{C}}, \mathcal{T}_\mathcal{I}$ 
    \State \textbf{Output:} initial activation set $\mathbf{S}_{0}$ 
    \Procedure{EnumInit}{$\mathcal{T}_{\mathcal{X}\setminus\mathcal{C}}, \mathcal{T}_\mathcal{I}$} 
    \For{$\hat{x}_{\mathcal{X}\setminus\mathcal{C}}\in \mathcal{T}_{\mathcal{X}\setminus\mathcal{C}}$ and $\hat{x}_\mathcal{I}\in \mathcal{T}_\mathcal{I}$} \Comment{Loop over all pairs of samples}
    \State $x_{left}, x_{right} \gets \hat{x}_{\mathcal{X}\setminus\mathcal{C}}, \hat{x}_\mathcal{I}$ 
    \State $x_{mid} = 0.5(x_{left} + x_{right})$ 
    \While{$x_{left} < x_{right}$} 
        \If{$\text{BoundaryLP}(x_{mid})$} \Comment{Check if the BoundaryLP is feasible}
            \State \textbf{Return} $\mathbf{S}(x_{mid})$ \Comment{Return $\mathbf{S}(x_{mid})$ as $\mathbf{S}_0$}
        \EndIf
        \If{$\overline{W}(\mathbf{S}(x^{\prime})) x_{mid} + \overline{r}(\mathbf{S}(x^{\prime})) \geq 0$} \Comment{Check if $b(x_{mid})>0$}
            \State $x_{right} \gets x_{mid}$ \Comment{Update the right point} 
        \Else 
            \State $x_{left} \gets x_{mid}$ \Comment{Update the left point}
        \EndIf
    \EndWhile
    \EndFor
    \EndProcedure
\end{algorithmic}
\label{alg:identify}
\end{algorithm}

\sysname utilizes NBFS for enumerating all activation sets along the zero-level set in a breadth-first search manner, starting from an initial set $\mathbf{S}_0$. The algorithm initializes a queue $\mathcal{Q}$ and a set $\mathcal{S}$ with $\mathbf{S}_0$. While $\mathcal{Q}$ is not empty, it dequeues an activation set $\mathbf{S}$ and checks if the set $\mathbf{S}\in\mathcal{B}$ by solving the boundary linear program $\text{BoundaryLP}(\mathbf{S})$. If so, $\mathbf{S}$ is identified and added to $\mathcal{S}$. The algorithm then explores its neighboring activation sets by flipping each neuron activation $(i, j)$ in $\mathbf{S}$. For each flip, it solves the unstable neuron linear program $\text{USLP}(\mathbf{S}, (i, j))$. If $\text{USLP}$ is feasible, the new activation set $\mathbf{S}'$ obtained by flipping $(i, j)$ is added to $\mathcal{Q}$ for further search on its neighbors. This process continues until all relevant activation sets are explored, resulting in a set $\mathcal{S}$ that contains activation sets potentially on the boundary.
\begin{algorithm}[h]
\caption{Enumerate Activated Sets}
\begin{algorithmic}[1]
    \State \textbf{Input:} $\mathbf{S}_0$ 
    \State \textbf{Output:} Set of activation sets $\mathcal{S}$ 
    \Procedure{NBFS}{$\mathbf{S}_0$} 
    \State Initialize queue $\mathcal{Q}$ with initial activation set $\mathbf{S}_0$ 
    \State Initialize sets $\mathcal{S}$ with $\mathbf{S}_0$ 
    \While{queue $\mathcal{Q}$ is not empty} 
        \State Dequeue $\mathbf{S}$ from $\mathcal{Q}$ \Comment{Dequeue the first activation set from the queue}
        \If{$\text{BoundaryLP}(\mathbf{S})$ is feasible} \Comment{Check if $\mathbf{S}$ is on a boundary activation set}
            \If{$\mathbf{S} \notin \mathcal{S}$} \Comment{If $\mathbf{S}$ is not already in $\mathcal{S}$}
                \State Add $\mathbf{S}$ to $\mathcal{S}$ 
            \EndIf
            \For{$(i,j) \in \{1,\ldots,L\} \times \{1,\ldots,M_{i}\}$} \Comment{For activation $(i,j)$ of each neurons}
                \If{$\text{USLP}(\mathbf{S}, (i,j))$} \Comment{Check if the neighbor may contain zero-level set}
                    \State Create $\mathbf{S}^{\prime}$ by flipping activation $(i,j)$
                    \State Add $\mathbf{S}^{\prime}$ to $\mathcal{Q}$ \Comment{Add adjacent activation set $\mathbf{S}^{\prime}$ to the queue}
                \EndIf
            \EndFor
        \EndIf
    \EndWhile
    \State \textbf{Return} $\mathcal{S}$ 
    \EndProcedure
\end{algorithmic}
\label{alg:NBFS}
\end{algorithm}

\begin{algorithm}[h]
\caption{Enumerate Hinges}
\begin{algorithmic}[1]
    \State \textbf{Input:} $\mathcal{S}$, $n$ 
    \State \textbf{Output:} $\mathcal{V}$ 
    \Procedure{HingeEnum}{$\mathcal{S}, n$} 
    \State Initialize hinge\_list as an empty list 
    \For{$k$ from $2$ to $n$} \Comment{Outer loop to iterate over combination sizes}
        \For{$\mathbf{S}\in \mathcal{S}$} \Comment{Iterate over all activation set}
            \If{$k=2$}
                \State $\mathcal{V}_c\gets\mathcal{N}_{\mathbf{S}}(\mathcal{S})$ 
            \Else
                \State $\mathcal{V}_c\gets\mathcal{V}^{(k-1)}$
            \EndIf
            \For{each combination $\mathcal{S}_c\in\mathcal{V}_c$ and $\mathbf{S}\notin \mathcal{S}_c$} \Comment{Iterate over combinations} 
                \If{$\mathbf{S}$ and $\mathcal{S}_c$ are not adjacent}
                    \State Continue \Comment{skip non-adjacent combinations}
                \EndIf
                \If{$\text{HingeLP}(\mathcal{S}_c\cup\{\mathbf{S}\})$} 
                    \State Add $\mathbf{V}\gets\mathcal{S}_c\cup\{\mathbf{S}\}$ to $\mathcal{V}^{(k)}$ \Comment{Add to set of corresponding $k$}
                \EndIf
            \EndFor
        \EndFor
        \State Add $\mathcal{V}^{(k)}$ to $\mathcal{V}$
    \EndFor
    \State \textbf{return} hinge\_list 
    \EndProcedure
\end{algorithmic}
\label{alg:hinge_search}
\end{algorithm}
\sysname enumerates all hinges $\mathbf{V}\in \mathcal{V}$ with Algorithm \ref{alg:hinge_search}. Algorithm~\ref{alg:hinge_search} outlines a method to enumerate all feasible hinge hyperplanes formed by combinations of activation sets up to size $n$. The algorithm takes as input a set of activation sets $\mathcal{S}$ and a maximum combination size $n$. It initializes an empty list to store feasible hinges. For each combination size $k$ from $2$ to $n$, the algorithm iterates over all activation sets in $\mathcal{S}$. For each activation set $\mathbf{S}$, it generates candidate combinations $\mathcal{V}_c$ based on adjacency—either the set of adjacent activation sets when $k=2$, or the feasible combinations from the previous iteration when $k > 2$. It then checks each candidate combination $\mathcal{S}_c$ by verifying adjacency and solving the hinge linear program $\text{HingeLP}(\mathcal{S}_c \cup \{\mathbf{S}\})$. If the $\text{HingeLP}$ is feasible, the combination is added to the list of feasible hinges $\mathcal{V}$. This process continues until all combinations up to size $n$ have been examined, resulting in a comprehensive list of feasible hinge hyperplanes that are essential for understanding the intersections of activation regions.

\subsection{Linear Programs for Enumeration}
\label{appendix:nlp_enum}


Given a state $x_{i}^{0}$, we define the activation set is $\mathbf{S}=\tau_S(x_{i}^{0})$. To determine if the activation set $\mathbf{S}\in \mathcal{B}$, we solve a linear program referred to as the boundary linear program. The program checks the existence of a state $x\in \overline{\mathcal{X}}(\mathbf{S}(x_{i}^{0}))$ that satisfies $\overline{W}(\mathbf{S}(x_{i}^{0}))^{T} x + \overline{r}(\mathbf{S}(x_{i}^{0})) = 0$. The boundary linear program ($\text{BoundaryLP}(\mathbf{S}(x_{i}^{0}))$) is defined as follows.
\begin{equation}
\label{eq:boundarylp}
    \text{BoundaryLP}(\mathbf{S}(x_{i}^{0})) = 
    \begin{cases}
    \text{find} & x \\
    \text{s.t.} & \overline{W}(\mathbf{S}(x_{i}^{0}))^{T} x + \overline{r}(\mathbf{S}(x_{i}^{0})) = 0 \\
                    & x \in \overline{\mathcal{X}}(\mathbf{S}(x_{i}^{0}))
    \end{cases}
\end{equation}
NBFS conducts its search in a breadth-first manner. To determine if a neighboring activation set, resulting from a flip in its $(i,j)$ neuron, may contain the boundary, NBFS solves a linear program, referred to as the unstable neuron linear program of $\text{USLP}(\mathbf{S}, (i,j))$. This linear program checks the existence of a state $x\in \overline{\mathcal{X}}(\mathbf{S})\cap\{x: W_{ij}x+r_{ij}=0\}$ that satisfies $\overline{W}(\mathbf{S}) x+\overline{r}(\mathbf{S})=0$. The unstable neuron linear program is defined as follows.
\begin{equation}
\label{eq:uslp}
    \text{USLP}(\mathbf{S}, (i,j)) = 
    \begin{cases}
    \text{find} & x \\
    \text{s.t.} & \overline{W}(\mathbf{S})^{T} x + \overline{r}(\mathbf{S}) = 0 \\
                    & W_{ij}(\mathbf{S})^{T} x + r_{ij}(\mathbf{S})=0 \\
                    & x \in \overline{\mathcal{X}}(\mathbf{S})
    \end{cases}
\end{equation}
Finally, we enumerate all hinges by solving the hinge linger program $\text{HingeLP}(\mathcal{S}_i)$, defined as follows. 
\begin{equation}
\label{eq:hingelp}
    \text{HingeLP}(\mathcal{N}^{(d)}_{\mathbf{S}}(\mathcal{S}_i)) = 
    \begin{cases}
    \text{find} & x \\
    \text{s.t.} & \overline{W}(\mathbf{S})^{T} x + \overline{r}(\mathbf{S}) = 0, \quad \forall \mathbf{S}\in \mathcal{N}_{\mathbf{S}}(\mathcal{S}_i)\\
                & x \in \mathcal{X}(\mathbf{S}), \quad \forall \mathbf{S}\in \mathcal{N}_{\mathbf{S}}(\mathcal{S}_i)
    \end{cases}
\end{equation}


\subsection{Nonlinear Programs for Verification}
\label{appendix:nlp}
The correctness condition (\eqref{eq:neural-containment}) can be verified for boundary hyperplane $\overline{\mathcal{X}}(\mathbf{S})$ by solving the  nonlinear program 
\begin{equation}
\label{eq:containment-verification}
\begin{array}{ll}
\min_x & h(x) \\
\mbox{s.t.} & \overline{W}(\mathbf{S})^{T}x + \overline{r}(\mathbf{S}) = 0, x \in \overline{\mathcal{X}}
(\mathbf{S})
\end{array}
\end{equation}
If we found some state $x$ such that $h(x)<0$ and $\overline{W}(\mathbf{S})^{T}x + \overline{r}(\mathbf{S}) = 0$, it indicates that the state lies on the boundary of the set $\mathcal{D}$ but not inside the set $\mathcal{C}$, i.e., $x\in \partial \mathcal{D} \notin \mathcal{C}$. This implies a violation of the condition $\mathcal{D}\subseteq\mathcal{C}$. 

Hyperplane counterexamples can be identified by solving the optimization problem
\begin{equation}
    \label{eq:hyperplane-verification-exact}
    \begin{array}{ll}
    \mbox{min}_{x} & \max{\{\overline{W}(\mathbf{S})^{T}(f(x)+g(x)u): u \in \mathcal{U}\}} \\
    \mbox{s.t.}& \overline{W}(\mathbf{S})^{T}x + \overline{r}(\mathbf{S}) = 0,  x \in \overline{\mathcal{X}}(\mathbf{S})
     \end{array}
\end{equation}
Consider a case $\mathcal{U} = \{D\omega : ||\omega||_{\infty} \leq 1\}$. The bounded input set $\mathcal{U}$ allows us to replace the maximization over $u$ with $L$-1 norm term $\|\overline{W}(\mathbf{S})^{T} g(x) D \|_1$, which simplifies the computational complexity. In this case, the problem reduces to the  nonlinear program 
\begin{equation}
\begin{split}
\label{eq:interval_minLf}
    \min_x  \quad & \overline{W}(\mathbf{S})^{T} f(x) + \|\overline{W}(\mathbf{S})^{T} g(x) D \|_1 \\
    \text{s.t.} \quad & \overline{W}(\mathbf{S})^{T} x + \overline{r}(\mathbf{S}) = 0,
                x \in \overline{\mathcal{X}}(\mathbf{S})
\end{split}
\end{equation}
If $\mathcal{U} = \mathbb{R}^{m}$, then  by \cite[Corollary 1]{zhang2023exact}, the problem can be reduced to the  nonlinear program 
\begin{equation}
\label{eq:single-set-nonlinear-prog-special-case}
\begin{array}{ll} 
\min_{x} & \overline{W}(\mathbf{S})^{T}f(x) \\
\mbox{s.t.} 
& \overline{W}(\mathbf{S})^{T}g(x) = 0, \overline{W}(\mathbf{S})x + \overline{r}(\mathbf{S}) = 0, x \in \overline{\mathcal{X}}
(\mathbf{S}) 
\end{array}
\end{equation}

The hinge $\mathbf{V} = \{\mathbf{S}_{1},\ldots,\mathbf{S}_{r}\}$ can be certified by solving the nonlinear optimization problem
\begin{equation}
    \label{eq:hinge-certify}
    \begin{array}{ll}
    \mbox{min}_{x} & \max{\{\overline{W}(\mathbf{S}_{l})^{T}(f(x)+g(x)u) : l=1,\ldots,r, u \mbox{ satisfies \eqref{eq:safety-condition-1}--\eqref{eq:safety-condition-1a}}\}} \\
    \mbox{s.t.} & x \in \overline{\mathcal{X}}(\mathbf{S}_{1}) \cap \cdots \cap \overline{\mathcal{X}}(\mathbf{S}_{r}), \overline{W}(\mathbf{S}_{1})^{T}x + \overline{r}(\mathbf{S}_{1}) = 0
    \end{array}
\end{equation}

\subsection{Proof of Theorem \ref{th:verification_guarantee}}
\label{appendix:veri_guarantee}
\begin{proof}
    By Lemma \ref{lemma:enu_init}, the initial boundary activation set $\mathbf{S}_0$ is ensured to be identified by the verification of \sysname in Line 4 Algorithm. \ref{alg:EEV}. 
    Given $\mathbf{S}_0$ and the enumeration in Line 5 and 6 Algorithm. \ref{alg:EEV} being complete, the completeness of $\mathcal{S}$ and $\mathcal{V}$ is guaranteed by Proposition \ref{prop:enumeration-complete}. 
    By dReal solving the equivalent NLPs, the conditions of Proposition \ref{prop:safety-condition}, are satisfied. Therefore, $b_{\theta}(x)$ is a valid NCBF
\end{proof}

\section{Experiments}
\subsection{Experiment Settings}
\label{appendix:exp_settings}
\textbf{Experiment Settings of Darboux: }
The dynamic model of Darboux is given as follows. 
\begin{equation}
    \left[\begin{array}{c}
    \dot{x}_1 \\
    \dot{x}_2
    \end{array}\right]=\left[\begin{array}{c}
    x_2+2 x_1 x_2 \\
    -x_1+2 x_1^2-x_2^2
    \end{array}\right].
\end{equation}
We define state space, initial region, and safe region as $\mathcal{X}:\left\{\mathbf{x} \in \mathbb{R}^2: x\in[-2,2]\times[-2,2] \right\}$, $\mathcal{I}:\left\{\mathbf{x} \in \mathbb{R}^2: 0 \leq x_1 \leq 1,1 \leq x_2 \leq 2\right\}$ and $\mathcal{C}:\left\{\mathbf{x} \in \mathbb{R}^2: x_1+x_2^2 \geq 0\right\}$ respectively. 

\textbf{Experiment Settings of the Obstacle Avoidance: }
The dynamic model of obstacle avoidance is given as follows. 
\begin{equation}
    \left[\begin{array}{c}
    \dot{x}_1 \\
    \dot{x}_2 \\
    \dot{\psi}
    \end{array}\right]=\left[\begin{array}{c}
    v \sin \psi \\
    v \cos \psi \\
    0
\end{array}\right]
+ \left[\begin{array}{c}
    0 \\
    0 \\
    u
\end{array}\right]. 
\end{equation}
We define the state space, initial region and safe region as $\mathcal{X}$, $\mathcal{I}$ and $\mathcal{C}$, respectively as 
\begin{equation}
    \begin{aligned}
    \mathcal{X}:&\left\{\mathbf{x} \in \mathbb{R}^3:x_1,x_2,\psi\in [-2,2]\times[-2,2]\times[-2,2] \right\} \\
    \mathcal{I}:&\left\{\mathbf{x} \in \mathbb{R}^3:-0.1 \leq x_1 \leq 0.1,-2 \leq x_2 \leq-1.8, \ -\pi/6<\psi<\pi / 6 \right\} \\
    \mathcal{C}:&\left\{\mathbf{x} \in \mathbb{R}^3: x_1^2+x_2^2 \geq 0.04\right\}
\end{aligned}
\end{equation}

\textbf{Experiment Settings of the Spacecraft Rendezvous: }
The state of the chaser is expressed relative to the target using linearized Clohessy–Wiltshire–Hill equations, with state $x=[p_x, p_y, p_z, v_x, v_y, v_z]^T$, control input $u=[u_x, u_y, u_z]^T$ and dynamics defined as follows. 
\begin{equation}
    \left[\begin{array}{c}
    \dot{p}_x \\
    \dot{p}_y \\
    \dot{p}_z \\
    \dot{v}_x \\
    \dot{v}_y \\
    \dot{v}_z
    \end{array}\right]=
    \left[\begin{array}{c c c c c c}
    1 &0 &0 &0 &0 &0 \\
    0 &1 &0 &0 &0 &0 \\
    0 &0 &1 &0 &0 &0 \\
    3n^2 &0 &0 &0 &2n &0 \\
    0 &0 &0 &-2n &0 &0 \\
    0 &0 &-n^2 &0 &0 &0 
    \end{array}\right]
    \left[\begin{array}{c}
    p_x \\
    p_y \\
    p_z \\
    v_x \\
    v_y \\
    v_z
    \end{array}\right]
+ \left[\begin{array}{c c c}
    0 &0 &0 \\
    0 &0 &0 \\
    0 &0 &0 \\
    1 &0 &0 \\
    0 &1 &0 \\
    0 &0 &1
\end{array}\right]
\left[\begin{array}{c}
    u_x \\
    u_y \\
    u_z 
    \end{array}\right]. 
\end{equation}
We define the state space and safe region as $\mathcal{X}$, initial safe region $\mathcal{X}_{\mathcal{I}}$ and $\mathcal{C}$, respectively as 
\begin{equation}
    \begin{aligned}
    \mathcal{X}:&\left\{\mathbf{x} \in \mathbb{R}^3: p, v, \in [-5,5]\times[-1,1]\right\} \\
    \mathcal{I}:&\left\{ r \geq 0.75\text{, where } r=\sqrt{p_x^2+ p_y^2+ p_z^2} \right\}\\
    \mathcal{C}:&\left\{ r \geq 0.25\text{, where } r=\sqrt{p_x^2+ p_y^2+ p_z^2} \right\}
\end{aligned}
\end{equation}

\textbf{hi-ord$_8$: } The dynamic model of hi-ord$_8$ is given as follows. 
\label{sup:hi-ord}
\begin{equation}
    x^{(8)} +20 x^{(7)}+170 x^{(6)}+800 x^{(5)}+2273 x^{(4)} +3980 x^{(3)}+4180 x^{(2)}+2400 x^{(1)}+576=0
\end{equation}
where we denote the $i$-th derivative of variable $x$ by $x^{(i)}$. 
We define the state space $\mathcal{X}$, initial region $\mathcal{X}_{\mathcal{I}}$ and safe region $\mathcal{C}$, respectively as 
\begin{equation}
\begin{aligned}
    \mathcal{X}:&\left\{x_1^2+\ldots+x_8^2 \leq 4\right\} \\
    \mathcal{I}:&\left\{(x_1-1)^2+\ldots+(x_8-1)^2 \leq 1\right\} \\
    \mathcal{C}:&\left\{\left(x_1+2\right)^2+\ldots+\left(x_8+2\right)^2 \geq 3\right\}
\end{aligned}
\end{equation}

\subsection{Synthesis Framework Evaluation}
\label{appendix:synthesis_framework}

The experimental results presented in Table~\ref{tab:ce_synthesis} demonstrate the effectiveness of Counter Example (CE) guided training on Darboux and hi-ord$_8$ system. In this method, after each training epoch, we calculate the Control Barrier Function (CBF) outputs on representative samples. If the CBF correctly categorizes the samples into safe and unsafe regions, the certification procedure is initiated. If the CBF fails certification, the counter example is added to the training dataset for retraining. Otherwise, training is stopped early.

We capped the maximum training epochs at 50 and conducted three rounds of training for each network structure and system using different random seeds. The results indicate that without CE, the training process could basrely generate a CBF that passes certification. In contrast, with CE enabled, there was a success rate of at least 1/3 for most network structure, with verifiable policies generated in as few as 10 epochs. This highlights the improvement in training efficiency and reliability with the incorporation of CEs.

\begin{table}[h]
\centering
\begin{tabular}{c|cc|c|cc}
\toprule
\multirow{2}{*}{Case}    & \multirow{2}{*}{$L$} & \multirow{2}{*}{$M$} & No CE & \multicolumn{2}{c}{With CE} \\ 
                         &                      &                      & sr    & sr        & min epoch       \\ \hline
\multirow{4}{*}{Darboux} & 2                    & 8                    & 0/3   & 3/3       & 38              \\ 
                         & 2                    & 16                   & 0/3   & 1/3       & 10              \\ 
                         & 4                    & 8                    & 0/3   & 1/3       & 43              \\
                         & 4                    & 16                   & 0/3   & 2/3       & 26              \\ \hline
\multirow{4}{*}{hi-ord$_8$}  & 2                    & 8                    & 1/3   & 1/3       & 15              \\
                         & 2                    & 16                   & 0/3   & 2/3       & 19              \\
                         & 4                    & 8                    & 0/3   & 0/3       & -               \\
                         & 4                    & 16                   & 0/3   & 2/3       & 13    \\
\bottomrule
\end{tabular}
\caption{Success rates (sr) and minimum epochs required for certification with and without Counter Example (CE) guided training for different network structures on Darboux and hi-ord$_8$ systems.}
\label{tab:ce_synthesis}
\end{table}

\subsection{Hyperparameters}
\label{appendix:hyper_para}

Table~\ref{tab:hyperparameter} shows values the following hyperparameters used during CBF synthesis:
\begin{itemize}
    \item $N_{\text{data}}$: number of samples to train CBF on.
    \item $a_1$: weight penalizing incorrect classification of safe samples in Equation~\ref{eq:corr_loss}.
    \item $a_2$: weight penalizing incorrect classification of unsafe samples Equation~\ref{eq:corr_loss}.
    \item $\lambda_{f}$: weight penalizing violation of Lie derivative condition of CBF in Equation~\ref{eq:uncons_opt}.
    \item $\lambda_{c}$: weight penalizing correct loss for in Equation~\ref{eq:uncons_opt}.
    \item $n_{\text{cluster}}$: number of clusters in $\mathcal{L}_{\mathcal{B}}$ regularization.
    \item $k_{\sigma}$: value of $k$ used in generalized sigmoid function to perform differentiable activation pattern approximation.
    \item $\epsilon_{\text{boundary}}$: the threshold for range-based approximation of CBF boundary.
\end{itemize}

\begin{table}[h]
\centering
\begin{tabular}{lllllllll}
\toprule
Case       & $N_{\text{data}}$ & $a_1$ & $a_2$ & $\lambda_{f}$ & $\lambda_{c}$ & $n_{\text{cluster}}$ & $k_{\sigma}$ & $\epsilon_{\text{boundary}}$ \\ \hline
Darboux    & 5000                                & 100   & 100   & 4.0           & 1.0           & N/A                  & N/A          & N/A                          \\ \hline
hi-ord$_8$ & 50000                               & 100   & 200   & 1.0           & 1.0           & N/A                  & N/A          & N/A                          \\ \hline
OA         & 10000                               & 100   & 100   & 2.0           & 1.0           & 5                    & 4            & 1.0                          \\ \hline
SR         & 10000                               & 100   & 100   & 2.0           & 1.0           & 5                    & 4            & 1.0                          \\ \bottomrule
\end{tabular}
\caption{Hyperparameters of CBF synthesis.}
\label{tab:hyperparameter}
\end{table}
\subsection{Sensitivity Analysis of Hyperparameters}
\label{appendix:sensitivity}

\begin{table}[h]
    \centering
    \begin{subtable}{0.31\linewidth}
        \centering
            \begin{tabular}{c|ccc}
            $\lambda_c$ & SR & ME & $N$ \\ \hline
            1   & 0/3 & x  & x  \\
            10  & 0/3 & x  & x  \\
            \textbf{100} & \textbf{3/3} & \textbf{17}   & \textbf{3265} \\
            200 & 3/3 & 17.7 & 2922 
            \end{tabular}
        \caption{Ablation study on $\lambda_c$}
        \label{tab:abl-lambda_c}
    \end{subtable}%
    \hfill
    \begin{subtable}{0.31\linewidth}
        \centering
            \begin{tabular}{c|ccc}
            $\lambda_f$ & SR & ME & $N$ \\ \hline
            1 & 3/3 & 16.3 & 3254 \\
            \textbf{2} & \textbf{3/3} & \textbf{17}   & \textbf{3265} \\
            4 & 3/3 & 17   & 3352 \\
            8 & 2/3 & 29.5 & 3419.5
            \end{tabular}
        \caption{Ablation study on $\lambda_f$}
        \label{tab:abl-lambda_f}
    \end{subtable}
    \hfill
    \begin{subtable}{0.31\linewidth}
        \centering
            \begin{tabular}{c|ccc}
            $k$ & SR & ME & $N$ \\ \hline
            1 & 3/3 & 18   & 3842 \\
            2 & 3/3 & 15.7 & 3523 \\
            \textbf{4} & \textbf{3/3} & \textbf{17}   & \textbf{3265} \\
            8 & 1/3 & 10   & 1984 
            \end{tabular}
        \caption{Ablation study on $k$}
        \label{tab:abl-sigmoid_k}
    \end{subtable}
    
    \caption{Ablation study for training hyperparameters. In each table, the bold lines indicate the baseline setting. \textbf{SR}: the success rate among runs with three random seeds. \textbf{ME}: the average first training epoch when a valid NCBF is obtained. $\mathbf{N}$: the average number of boundary hyperplanes.}
    \label{tab:ablation_study}
    \vspace{-0.2in}
\end{table}

Next, we performed a sensitivity analysis of the hyperparameters. We chose the case study of Spacecraft Rendezvous with the number of layers $L=4$ and the number of hidden units per layer $N=8$. We studied the sensitivity of the training performance to the hyperparameters $\lambda_{\mathcal{B}}$, $\lambda_{f}$, and $\lambda_{c}$ in Equation 8, corresponding to the weightings for regularizing the number of boundary hyperplanes, NCBF value violation, and NCBF Lie derivative violation, respectively. We also studied the sensitivity to the hyperparameter $k$ employed in the modified sigmoid function $\sigma_k (z)=\frac{1}{1+ \exp(-k \cdot z)}$ to approximate the regularization pattern. We compared against the settings used in the original paper: $\lambda_{\mathcal{B}}=10$, $\lambda_{f}=2$, $\lambda_{c}=100$, and $k=4$. For each hyperparameter, we chose four values to perform the ablation study: $\lambda_{\mathcal{B}} \in \{0, 1, 10, 50\}$, $\lambda_{f} \in \{1, 2, 4, 8\}$, $\lambda_{c} \in \{1, 10, 100, 200\}$, and $k \in \{1, 2, 4, 8\}$. For each setting, we performed three runs with different random seeds. We measured the results by \textbf{Success Rate (SR)}, \textbf{Min Epoch (ME)}, and $\mathbf{N}$, as described in the caption of Table~\ref{tab:ablation_study}.

$\lambda_c$ regularizes the shape of the NCBF by penalizing incorrectly categorized samples. Table~\ref{tab:abl-lambda_c} indicates that when $\lambda_c$ is too small, the training procedure fails to train an NCBF that correctly separates the safe and unsafe regions, resulting in failure of certification. Meanwhile, a larger weight delivers similarly good performance.

$\lambda_f$ penalizes violations of Lie derivative conditions. Table~\ref{tab:abl-lambda_f} shows that the result is not sensitive to this hyperparameter, as this term quickly goes down to 0 when the Lie derivative condition is satisfied. We note that over-penalizing this condition should be avoided since the NCBF would otherwise learn an unrecoverable incorrect shape, as demonstrated by the failure case when $\lambda_f=8$.

$\lambda_{\mathcal{B}}$ has been studied in the original paper, with detailed analysis in Section 5.2. The boundary regularization term reduces the number of boundary hyperplanes and benefits convergence.

Table~\ref{tab:abl-sigmoid_k} shows the importance of the term $k$ used in the modified sigmoid function. Since this term appears in the exponential part of the sigmoid function, when it is too large, it leads to gradient explosions during backpropagation, which crashes the training process. Conversely, a reasonably larger $k$ better approximates the activation pattern, leading to a reduced number of boundary hyperplanes.

In summary, balancing the hyperparameters is relatively straightforward, as the training performance remains robust across a wide range of hyperparameter values. When training failures do occur, we can systematically identify the cause from observation. This enables proper guidance in choosing and adjusting the appropriate hyperparameters.

\end{document}